\documentclass{jfm}

\usepackage{graphicx}
\usepackage{newtxtext}
\usepackage{newtxmath}
\usepackage{booktabs}
\usepackage{siunitx}

\usepackage{natbib}
\usepackage{hyperref}
\hypersetup{
    colorlinks = true,
    urlcolor   = blue,
    citecolor  = black,
}
\newcommand{\pd}{\partial}
\newcommand{\Nu}{\mbox{\textit{Nu}}}
\newcommand{\Gr}{\mbox{\textit{Gr}}}
\newcommand{\Ra}{\mbox{\textit{Ra}}}
\newcommand{\Ri}{\mbox{\textit{Ri}}}
\renewcommand{\Rey}{\mbox{\textit{Re}}}
\renewcommand{\Pran}{\mbox{\textit{Pr}}}

\newcommand{\bu}{\boldsymbol{u}}
\newcommand{\ubar}{\overline{u}}
\newcommand{\wbar}{\overline{w}}
\newcommand{\grad}{\boldsymbol{\nabla}}

\newcommand{\RomanNumeralCaps}[1]
\linenumbers

\graphicspath{{figs/}}

\title{Turbulent mixed convection in vertical and horizontal channels}
\shorttitle{Turbulent mixed convection in vertical and horizontal channels}

\author{
    Christopher J.~Howland\aff{1} \corresp{\email{\href{mailto:c.j.howland@outlook.com}{c.j.howland@outlook.com}}},
    Guru Sreevanshu Yerragolam\aff{1},
    Roberto Verzicco\aff{2,3,1}
    \and Detlef Lohse\aff{1,4}
}
\shortauthor{C.J.~Howland, G.S.~Yerragolam, R.~Verzicco and D.~Lohse}

\affiliation{
    \aff{1}Physics of Fluids Group, Max Planck Center for Complex Fluid Dynamics, and J.~M.~Burgers Centre for Fluid Dynamics, University of Twente, P.O.~Box 217, 7500AE Enschede, The Netherlands
    \aff{2}Dipartimento di Ingegneria Industriale, University of Rome `Tor Vergata', Roma 00133, Italy
    \aff{3}Gran Sasso Science Institute - Viale F. Crispi, 7 67100 L'Aquila, Italy
    \aff{4}Max Planck Institute for Dynamics and Self-Organization, 37077 Göttingen, Germany
}

\begin{document}
\maketitle

\begin{abstract}
Turbulent shear flows driven by a combination of a pressure gradient and buoyancy forcing are investigated using direct numerical simulations.
Specifically, we consider the setup of a differentially heated vertical channel subject to a Poiseuille-like horizontal pressure gradient.
We explore the response of the system to its three control parameters: the Grashof number $\Gr$, the Prandtl number $\Pran$, and the Reynolds number $\Rey$ of the pressure-driven flow.
From these input parameters, the relative strength of buoyancy driving to the pressure gradient can be quantified by the Richardson number $\Ri=\Gr/\Rey^2$.
We compare the response of the mixed vertical convection configuration to that of mixed Rayleigh--Bénard convection and find a nearly identical behaviour, including an increase in wall friction at higher $\Gr$ and a drop in the heat flux relative to natural convection for $\Ri=O(1)$.
This closely matched response is despite vastly different flow structures in the systems.
No large-scale organisation is visible in visualisations of mixed vertical convection -- an observation that is quantitatively confirmed by spectral analysis.
This analysis, combined with a statistical description of the wall heat flux, highlights how moderate shear suppresses the growth of small-scale plumes and reduces the likelihood of extreme events in the local wall heat flux.
Vice versa, starting from a pure shear flow, the addition of thermal driving enhances the drag due to the emission of thermal plumes.

\end{abstract}


\section{Introduction}
\label{sec:headings}

The transport of heat by turbulent convection is integral to a wide range of natural and engineering applications, from building ventilation to the atmospheric boundary layer and the near-surface ocean.
All of these examples can, under the right conditions, be classified as mixed convection.
Mixed convection describes the scenario where both buoyancy and shear forces are relevant to the dynamics.
This is in contrast to natural convection, where a flow is solely driven by density differences within the fluid, and forced convection, where buoyancy is negligible and the transport of heat is identical to that of a passive scalar.
The relative importance of buoyancy compared to the imposed shear  is quantified by the Richardson number $Ri$, with the extreme cases $Ri=\infty$ for purely thermal driving and $Ri=0$ for purely shear or pressure driving.

The foundational work on mixed convection by \citet{obukhov_turbulence_1946} was motivated by understanding the dynamics of the surface layer of the atmosphere.
\citeauthor{obukhov_turbulence_1946} supposed that the dynamics were solely determined by the surface friction (quantified by the friction velocity $u_\tau$), the surface heat flux $q$, and gravity $g$, such that dimensional analysis revealed a single possible length scale $L_O = u_\tau^3 / g\alpha q$ that could describe the system.
Using this length to rescale the problem, \citet{monin_basic_1954} derived what is now known as the Monin--Obukhov Similarity Theory (MOST) where `universal' functions of $z/L_O$ are used to describe the mean velocity and temperature profiles in stably or unstably stratified shear flows.
These universal functions are obtained by interpolating between the extreme cases of natural convection and forced convection, which were updated by \citet{kader_mean_1990} for unstable (i.e.~convecting) boundary layers.
A historical overview of MOST is provided in \citet{foken_50_2006} and the theory has been extremely popular in atmospheric and oceanic applications.
However, some of the assumptions underlying MOST have recently been coming under further scrutiny, particularly the power-law dependence of the mean profiles in the convective regime \citep{cheng_logarithmic_2021}.

Mixed convection has also often been studied in simple, canonical flow configurations where the system response is only dependent on a small number of dimensionless input parameters.
A popular approach has been to introduce horizontal forcing into the classical Rayleigh--Bénard setup, either through a horizontal pressure gradient (Poiseuille--Rayleigh--Bénard, P-RB) or by setting one of the boundary plates in motion (Couette--Rayleigh--Bénard, C-RB).
The linear stability of the P-RB system was studied by \citet{gage_stability_1968} who found that streamwise perturbations are suppressed by the introduction of shear, so the fastest growing mode takes the form of convective rolls in the plane orthogonal to the mean flow.
Note that the critical Rayleigh number $\Ra_c=1708$ and the fastest growing wavelength $\lambda=2\sqrt{2}H$ do not change compared to natural RB since the linear problem is unchanged in the orthogonal plane.
Such streamwise-aligned rolls were observed experimentally by \citet{richter_interaction_1975-1} in C-RB although, since their setup was motivated by mantle convection, their working fluid had a very high Prandtl number of $\Pran=8600$ and low Reynolds numbers.
\citet{domaradzki_direct_1988} performed simulations of C-RB, also finding organisation into streamwise-aligned rolls but at the largest wavelength of the domain.
More recent direct numerical simulations in larger domains by \citet{pirozzoli_mixed_2017} for P-RB and \citet{blass_flow_2020,blass_effect_2021} for C-RB highlight how these large scale structures contribute a large proportion of the heat and momentum flux in mixed RB, and how their wavelength depends on the Richardson number $\Ri$ of the system.
\citet{madhusudanan_navier-stokes--based_2022} recently reproduced the wide rolls using a linear model coupled to eddy diffusivities, showing that they are primarily generated through a classical lift-up mechanism.

The response of canonical mixed convection systems can be quantified using the friction coefficient $C_f$ and the Nusselt number $\Nu$, which measure the dimensionless skin friction and heat flux.
In forced convection, where buoyancy is negligible, both Poiseuille and Couette flows exhibit an identical response in $C_f$ when appropriately scaled using the centreline velocity \citep{orlandi_poiseuille_2015}.
\citet{scagliarini_law_2015} observed an increase in the streamwise friction coefficient in P-RB relative to pure Poiseuille flow, for which they proposed a modified formulation of the log-law for the mean velocity in the presence of convection.
An intriguing phenomenon of mixed RB is found in the response of the heat flux, which varies non-monotonically with Reynolds number $\Rey$ for a fixed Rayleigh number and Prandtl number.
$\Nu$ first decreases relative to the natural convection case before increasing at high $\Rey$ as the flow enters the forced convection regime \citep{blass_flow_2020,blass_effect_2021}.
This behaviour, not predicted by MOST, has been attributed to the sweeping away of thermal plumes by the imposed horizontal flow.
The plume sweeping concept has since been applied to form phenomonological models of the system \citep{scagliarini_heat-flux_2014}.
Similar to the response of the friction coefficient, an identical response is found in P-RB and C-RB when appropriately scaled and the decrease in $\Nu$ has recently been shown to collapse onto a single curve when the Reynolds number of the shear flow is considered relative to the Reynolds number of the natural convection \citep{yerragolam_scaling_2024}.
\citet{yerragolam_scaling_2024} also provide a theoretical estimate for this decrease in heat flux based on an extension of the \citet{grossmann_scaling_2000,grossmann_thermal_2001} theory for RB convection to mixed RB.

The interplay of shear and convection plays an important role in another canonical natural convection problem: the differentially heated vertical channel, often simply referred to as vertical convection (VC).
In this configuration, convection drives flow parallel to the boundary plates, generating a mean shear at the walls and in the bulk of the channel.
An analogy can be drawn between the large-scale circulation in RB and the vertical mean flow in VC, but since the vertical buoyancy flux is not equivalent to the heat flux of interest in VC, the \citet{grossmann_scaling_2000} approach of linking heat flux and kinetic energy dissipation cannot be directly applied.
Nevertheless, \citet{ng_vertical_2015} found similar scaling relations to RB for heat flux and dissipation rates in VC when conditionally sampling either the boundary layers or the bulk.
Recent simulations at varying Prandtl number \citep{howland_boundary_2022} have prompted renewed efforts to understand the boundary layer theory limiting the global response of the system \citep{ching_heat_2023} and the dynamics setting the mean profiles in the channel centre \citep{li_mean_2023}.

Mixed convection in a vertical channel, where an additional pressure-driven forcing is applied to the VC configuration, has been less well studied than mixed RB.
The majority of studies into these flows \citep[e.g.][]{kasagi_direct_1997,wetzel_buoyancy-induced_2019,guo_direct_2022} impose a mean pressure gradient in the vertical direction that directly opposes or enhances the mean flow due to convection.
Although this configuration is relevant to some industrial applications, from a physical perspective it breaks the symmetry of the channel, with the boundary layers at each wall subject to different shear stresses.
In this study, we instead impose a horizontal pressure gradient in the channel, which leads to symmetric profiles of horizontal velocity and higher order statistics while retaining the anti-symmetric profiles of mean vertical velocity and temperature from VC.
Although we approach this configuration from a fundamental point of view, the crossflow setup can be relevant to industrial heat exchangers in a wide range of applications.
We are only aware of one other paper discussing such a system \citep{el-samni_direct_2005}, which highlights tilted structures at the wall and a modification of the near-wall Reynolds stresses.
However, the results of \citet{el-samni_direct_2005} are mainly descriptive and cover a limited parameter range.

In the current manuscript, we use direct numerical simulations to explore the dynamics of turbulent mixed convection in a vertical channel for a wide range of parameters, focusing on the transition between natural convection and forced convection.
The manuscript is organised as follows.
Firstly, in \S\ref{sec:setup} we describe the problem setup and details of the numerical simulations, before presenting visualisations of the resulting flow in \S\ref{sec:flow_viz}.
The global response of the system is investigated in terms of the friction coefficients and the Nusselt number, and compared with the mixed Rayleigh--Bénard flow in \S\ref{sec:global_response}.
We then turn to wall-normal profiles in mixed VC in \S\ref{sec:profiles}, primarily focusing on the balances in the mean momentum budgets.
Detailed analysis of the heat transport is then performed by spectral analysis in \S\ref{sec:spectra} and through statistics of the boundary layers in \S\ref{sec:BL_stats}.
Finally, our conclusions are presented in \S\ref{sec:conclusion}, where we discuss the implications of our results and provide an outlook on future research in mixed convection.

\section {Simulation setup and numerical methods \label{sec:setup}}

\begin{figure}
    \centering
    \includegraphics[width=\textwidth]{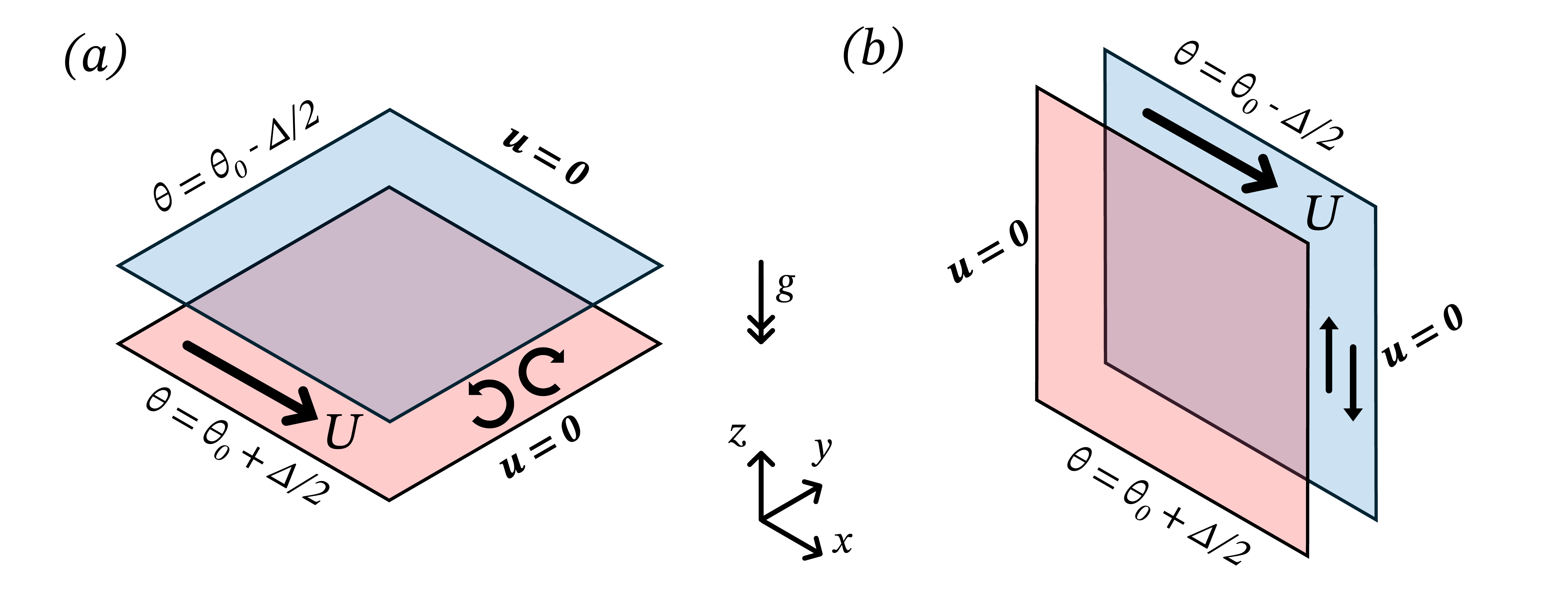}
    \caption{
        Schematic of canonical mixed convection systems.
        $(a)$ Mixed Rayleigh--Bénard convection with Poiseuille-type forcing as studied by e.g.~\citet{domaradzki_direct_1988,pirozzoli_mixed_2017,yerragolam_scaling_2024}.
        $(b)$ Mixed vertical convection with Poiseuille-type forcing, as used for the new simulations presented in this manuscript.
        Large arrows highlight the imposed horizontal pressure gradient, and smaller arrows highlight the large-scale flows driven by buoyancy in each configuration.
        The acceleration due to gravity is highlighted by a double arrow.
    }
    \label{fig:schematic}
\end{figure}

We perform numerical simulations of the flow arising in a fluid confined between two parallel no-slip vertical walls.
The walls are separated by a horizontal distance $H$ and are held at fixed temperatures, with a temperature difference of $\Delta$ between the plates.
As in the schematic in figure \ref{fig:schematic}\textit{(b)}, we take the $x$-coordinate to be horizontal and parallel to the plates, the $y$-coordinate to be normal to the boundaries, and $z$ to be in the vertical direction.
We consider a fluid satisfying the Oberbeck--Boussinesq approximation, such that changes in density are only significant in the buoyancy term and are linear with respect to changes in temperature.
We therefore treat the velocity field $\bu=(u,v,w)$ as divergence-free ($\grad \cdot \bu = 0$) and satisfying the Navier--Stokes momentum equation
\begin{equation}
    \pd_t \bu + \left(\bu \cdot \grad\right) \bu = -\rho^{-1} \grad p + \nu \nabla^2 \bu + g \alpha \theta \boldsymbol{\hat{z}} + G \boldsymbol{\hat{x}}, \label{eq:NSmom}
\end{equation}
where $\rho$ is the mean fluid density (assumed constant), $p$ is the pressure, $\nu$ is the kinematic viscosity, $g$ is the acceleration due to gravity, and $\alpha$ is the thermal expansion coefficient.
A time-dependent, spatially-uniform forcing $G(t)$ is applied in the streamwise ($x$) direction to maintain a constant mean flow $\langle u \rangle = U$.
The magnitude of this forcing is computed at every time step to exactly cancel out any variation in the mean flow.
Previous work has shown that such a forcing produces near-identical results to those of a constant pressure gradient \citep{quadrio_does_2016}, but allows us to use the mean flow strength as an input parameter.
The temperature field $\theta$ satisfies the advection-diffusion equation
\begin{equation}
    \pd_t \theta + \left(\bu \cdot \grad\right) \theta = \kappa \nabla^2 \theta , \label{eq:temp_evo}
\end{equation}
where $\kappa$ is the thermal diffusivity.
Periodic boundary conditions are applied to both $\bu$ and $\theta$ in the $x$ and $z$ directions.
Unless otherwise stated, the aspect ratio of the domain in these periodic directions is taken as $\Gamma=L/H=8$, such that the length of the domain $L$ in $x$ and $z$ is equal and eight times the plate separation distance $H$.

We perform direct numerical simulations of equations \eqref{eq:NSmom} and \eqref{eq:temp_evo} using a highly-parallelised flow solver that computes spatial derivatives using second-order central finite differences on a staggered grid configuration.
The wall-normal diffusive terms are evolved in time using a Crank--Nicolson scheme and all other terms are treated explicitly using a three-stage Runge--Kutta method.
An adaptive time step is chosen using the constraint of a maximum Courant number of 1.
The velocity is kept divergence-free to machine precision using a pressure correction step at each time step that is implemented with fast Fourier transforms in the periodic directions and a tridiagonal matrix solver in the wall-normal direction.
A multiple-resolution technique is applied to evolve the velocity and temperature fields on independent grids, with cubic Hermite interpolation used for the buoyancy forcing and the advection of temperature.
Detailed overviews of the numerical discretization, the domain decomposition strategy and the multiple-resolution technique can be found in \citet{verzicco_finite-difference_1996,van_der_poel_pencil_2015,ostilla-monico_multiple-resolution_2015} as well as in our software documentation.

The physical input parameters of the system are the Rayleigh number, the Prandtl number, and the Reynolds number
\begin{align}
    \Ra&= \frac{g\alpha \Delta H^3}{\nu \kappa}, &
    \Pran&= \frac{\nu}{\kappa},&
    \Rey&=\frac{U H}{\nu}. \label{eq:RaPrRe_def}
\end{align}
When considering the strength of the flow driven by convection, it is often useful to consider the Grashof number as the relevant input parameter, and when comparing the relative strengths of buoyancy to pressure driving, we can construct a Richardson number as
\begin{align}
    \Gr = \frac{\Ra}{\Pran} &= \frac{g \alpha \Delta H^3}{\nu^2} = \left(\frac{U_f H}{\nu}\right)^2, &
    \Ri = \frac{\Gr}{\Rey^2} &= \frac{g \alpha \Delta H}{U^2} = \left(\frac{U_f}{U}\right)^2 . \label{eq:GrRi_def}
\end{align}
These can both be considered as input parameters.
Above we have written $U_f = \sqrt{g\alpha \Delta H}$ as the free-fall velocity scale to give insight on the interpretation of these parameters.

In this study, we perform two sets of simulations to compare the relative impacts of the various input parameters.
Firstly, we fix $\Gr=10^6$ and vary $1\leq \Pran \leq 10$ along with $10^{2.5} \leq \Rey \leq 10^4$, which correspond to Richardson numbers of $10^{-2} \leq \Ri \leq 10$.
For the second set, we fix $\Pran=1$ and increase $\Gr$ up to $10^8$ while again varying $\Rey$ up to $10^4$.
A detailed overview of the parameters used for each simulation are given in table \ref{tab:params} of appendix \ref{app:numerics}.
Each simulation is run to a statistically steady state, in which the flow statistics are computed and averaged for a minimum of 200 advective time units.
For high $\Ri\geq 1$ the relevant time unit is $H/U_f$, whereas for low $\Ri\leq 1$ the relevant time unit is $H/U$.
The wall-normal grid spacing is stretched following \citet{pirozzoli_natural_2021} to ensure sufficient resolution close to the wall, such that $\Delta y^+ < 0.1$ at the wall for the base velocity grid.
In the periodic directions, the uniform grid spacing satisfies $\Delta x^+ \leq 5.4$ in every simulation.
At the centre of the domain, the spacing of the refined grid satisfies $\Delta y_r \leq 1.05 l_B$, $\Delta x_r < 1.4 l_B$, where $l_B$ is the Batchelor scale computed using the plane-averaged TKE dissipation rate over the mid-plane.
Full details of the grid sizes are provided in appendix \ref{app:numerics}.

\section{Flow visualisation \label{sec:flow_viz}}

\begin{figure}
    \centering
    \includegraphics{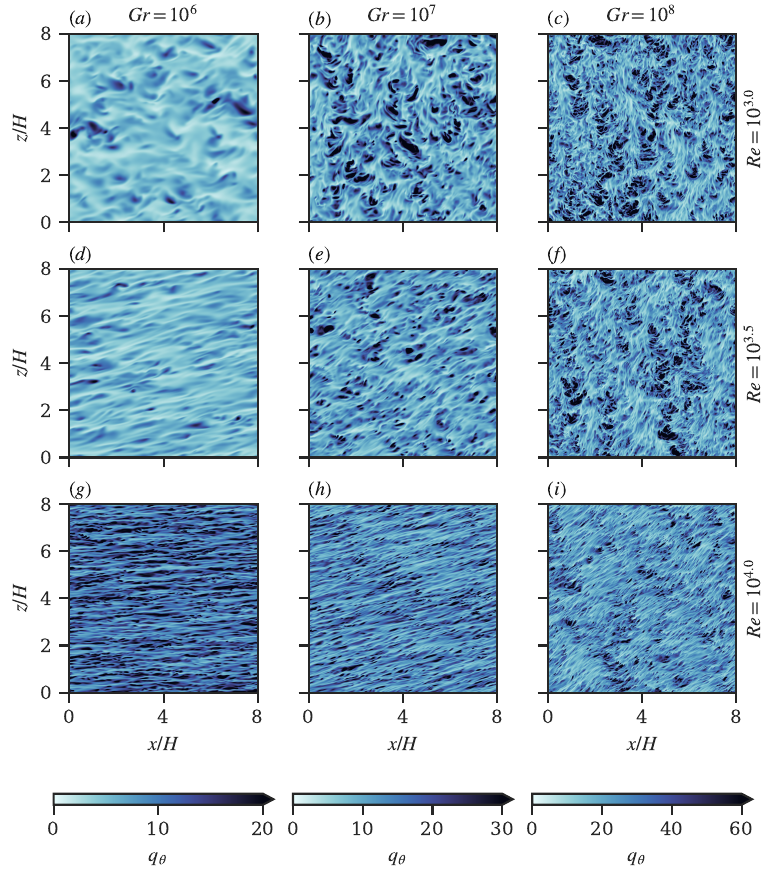}
    \caption{
        Vertical planes of the instantaneous local wall-normal heat flux at the boundary $y=0$.
        All simulations shown are with fixed $\Pran=1$.
        The Richardson number $Ri=Gr/\Rey^2$ is comparable along diagonals from the upper left to the lower right, with the largest value ($Ri=100$) in the top right and the lowest value ($Ri=0.01$) in the bottom left.
        This figure is also available as an \href{https://cocalc.com/share/public_paths/c690a4a6317a519769e8c2b6c3e5c97a40c2338f/figure\%202/figure\%202.ipynb}{interactive JFM notebook}.
    }
    \label{fig:wall_planes}
\end{figure}

We begin with a qualitative comparison of the simulations through visualisations of the temperature field and the local heat flux.
Figure \ref{fig:wall_planes} displays the instantaneous local wall-normal heat flux at the boundary plate $y=0$ for cases with fixed $\Pran=1$, and a range of $10^6\leq Gr\leq 10^8$, $10^3\leq \Rey\leq 10^4$.
The dimensionless heat flux plotted here is defined as
\begin{equation}
    q_\theta(x,z,t) = -\frac{H}{\Delta} \left. \frac{\pd \theta}{\pd y}\right|_{y=0}, \label{eq:heat_flux_def}
\end{equation}
such that its time- and plane-averaged value is equivalent to the Nusselt number, $\Nu=\overline{q_\theta}$.

\begin{figure}
    \centering
    \includegraphics{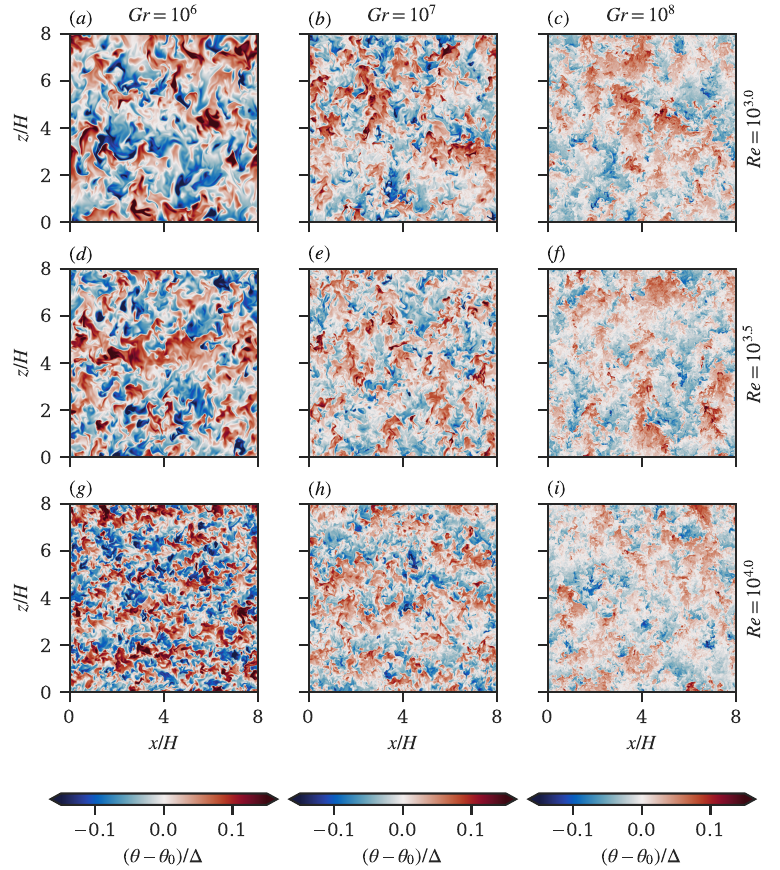}
    \caption{
        Vertical planes of the temperature field at the channel centre $y=H/2$.
        See the caption of figure \ref{fig:wall_planes} for details of the simulations presented.
        $\theta_0$ is the arbitrary reference temperature that is the midpoint of the two boundary values.
        This figure is also available as an \href{https://cocalc.com/share/public_paths/c690a4a6317a519769e8c2b6c3e5c97a40c2338f/figure\%203/figure\%203.ipynb}{interactive JFM notebook}.
    }
    \label{fig:midplanes}
\end{figure}

As mentioned above, the relative strength of convection to the horizontal flow can be characterised by the Richardson number $Ri=Gr/\Rey^2$, which is constant along diagonals in figure \ref{fig:wall_planes}.
At high $Ri$, as in panel $(c)$, the horizontal flow has little impact on the local distribution of the wall heat flux.
The near-wall temperature structures are the same as the case in the absence of a crossflow, with regions of large local heat flux (dark spots) separated by longer, streaky structures aligned in the vertical \citep{howland_boundary_2022}.
As $Ri$ decreases, such as in panels $(e)$ and $(i)$ where $Ri=1$, the prominence of the large heat flux regions diminishes and the streaks become aligned in the diagonal.
This visualises the local direction of the flow along the wall, which at $Ri=1$ is due to a combination of the vertical convection and the horizontal pressure gradient.
At lower $Ri$, these structures become more aligned with the horizontal, eventually spanning the domain as in panel $(g)$, which is reminiscent of classical low-speed streaks in turbulent channel flow \citep{kline_structure_1967,antonia_analogy_2009}.
A more quantitative analysis of the change in the near-wall heat flux distribution will be provided in \S\ref{sec:BL_stats}.

In figure \ref{fig:midplanes}, we present visualisations of the same simulations but now at the midplane of the simulation domain $y=H/2$.
As would be expected, the fields at higher $\Gr$ and $\Rey$ exhibit structures with a wider range of spatial scales.
Aside from this dynamical range, the effect of increasing $\Rey$ is less noticeable at the midplane than at the wall in figure \ref{fig:wall_planes}.
At the centre of a turbulent channel, the mean profile of horizontal velocity is relatively flat, with zero mean shear and a local minimum in turbulent kinetic energy.
By contrast in vertical convection, a mean shear in the vertical velocity drives the generation of turbulence in the bulk, argued by \citet{li_mean_2023} to follow a mixing layer-like behaviour.
In figure \ref{fig:midplanes}, greater mixing at higher $\Gr$ leads to smaller values of the temperature perturbations in the midplane, although the same effect is not evident as $\Rey$ increases.
Compared to the mixed Rayleigh--Bénard system, where gravity is orthogonal to the plates and the temperature field organises into large, streamwise-aligned coherent rolls, the fields in mixed VC appear rather featureless.
A hint of such large-scale rolls is only noticeable for the cases dominated by strong pressure driving with high $\Rey$ and low $\Ri$ (panels $(g,h)$) where the temperature structures appear more aligned along the streamwise ($x$) axis.
This contrast in the organisation of mixed convection systems will be investigated more quantitatively in \S\ref{sec:spectra} through spectral analysis, but we first turn to the global responses of the two mixed convection systems in the next section. 

\section{Global response quantities in mixed convection systems \label{sec:global_response}}
 
In this section, we compare the responses of the mixed vertical convection setup with those of mixed Rayleigh--Bénard using the simulation data of \citet{yerragolam_scaling_2024}.
Those simulations cover a comparable range of parameters of $10^6\leq Gr \leq 10^8$, $0.5\leq Pr \leq 5$, and $Re\leq 10^4$ in a domain of streamwise aspect ratio $\Gamma_x=8$ and spanwise aspect ratio $\Gamma_y=4$.
The flow solver shares an identical code base except for the multiple-resolution technique, which is only used in the newly reported mixed VC simulations.

\subsection{Friction coefficients}

A key global response parameter in shear-driven flows is the friction coefficient
\begin{equation}
    C_f = \frac{2 \tau}{\rho U^2} , \label{eq:Cf_def}
\end{equation}
where $\tau$ is the wall shear stress and $U$ is the velocity magnitude in the bulk.
Since the friction coefficient is solely determined by the velocity field, we expect the dependence on the Prandtl number to be relatively weak, and consider a relationship $C_f(\Rey)$.
Power-law scalings for $C_f(\Rey)$ can be derived for laminar boundary layer flows, with for example $C_f\sim \Rey^{-1}$ applicable to Couette or Poiseuille flows and $C_f \sim \Rey^{-1/2}$ arising from the classical Blasius boundary layer solution \citep{schlichting_boundary-layer_2016}.
For a turbulent boundary layer in the sense of Prandtl and von Kármán, the friction coefficient satisfies the relation
\begin{equation}
    \sqrt{\frac{2}{C_f}} = \frac{1}{\kappa_u} \log{\left( \Rey  \sqrt{\frac{C_f}{8}} \right) } + B, \label{eq:Cf_Prandtl}
\end{equation}
known as the Prandtl friction law after \citet{prandtl_zur_1932}.
The von Kármán constant $\kappa_u$ typically takes a value around 0.4 and the intercept $B$ close to 4 but the exact values, their universality and the way in which they are fit to data remain an active topic of research \citep{monkewitz_hunt_2023}.
Due to our similar setup and numerical methods, we take the values suggested by \citet{pirozzoli_turbulence_2014} of $\kappa_u = 0.41$ and $B=5$.

\begin{figure}
    \centering
    \includegraphics{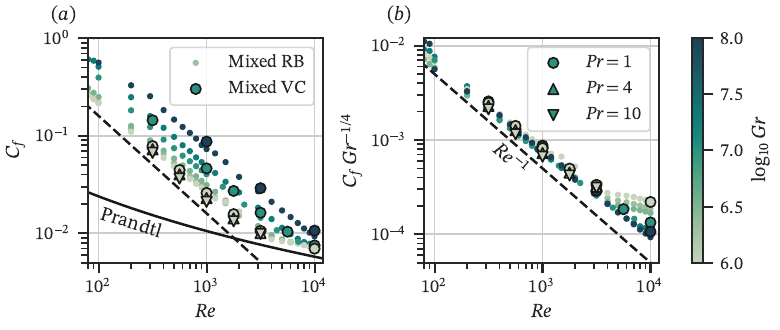}
    \caption{
        $(a)$ Friction coefficients calculated for the streamwise component of velocity as a function of Reynolds number.
        $(b)$ Friction coefficients normalised by $Gr^{1/4}$ to collapse the data at low $\Rey$.
        Large markers are used for the mixed vertical convection simulations, whereas small dots show the data from the mixed Rayleigh--Bénard cases of \citet{yerragolam_scaling_2024}.
        Grashof numbers are highlighted by the colour of the markers, and for the mixed VC cases different markers signify different $\Pran$.
        Black dashed lines show a scaling law of $\Rey^{-1}$, and the black solid line marks the Prandtl friction law of \eqref{eq:Cf_Prandtl}.
        This figure is also available as an \href{https://cocalc.com/share/public_paths/c690a4a6317a519769e8c2b6c3e5c97a40c2338f/figure\%204/figure\%204.ipynb}{interactive JFM notebook}.
    }
    \label{fig:Cf_Re_horiz}
\end{figure}

Since the mixed VC flow is driven in orthogonal directions by the pressure gradient and by buoyancy, we can construct separate friction coefficients for each component of the wall shear stress.
Understanding the response of the friction coefficients in this context relies on choosing an appropriate Reynolds number for each component of the flow.
For the streamwise ($x$) direction in which the mean flow is imposed by a pressure gradient, this Reynolds number is simply the input parameter defined in \eqref{eq:RaPrRe_def}.

We consider the response of the streamwise friction coefficient in figure \ref{fig:Cf_Re_horiz}, where only the streamwise component of the shear stress ${\tau=\rho\nu \pd_y \overline{u}}$ is applied to the definition \eqref{eq:Cf_def}.
The global response of mixed VC is near-identical to that of mixed RB, with a transition from a laminar power-law scaling to the Prandtl friction law of \eqref{eq:Cf_Prandtl}.
For comparable parameter values, the largest difference in $C_f$ between the mixed RB and mixed VC cases is 16\%.
In the laminar scaling regime, stronger buoyancy driving (characterised by higher $\Gr$) leads to an increase in the streamwise skin friction for a given $\Rey$.
As suggested earlier, the dependence of $C_f$ on $Pr$ is very weak compared to the other control parameters.
Unlike in standard Poiseuille or Couette flow, where a subcritical transition arises due to instability of the laminar base flow and leads to a jump in $C_f$, the streamwise boundary layer transition in mixed convection systems appears smooth.
We anticipate that the laminar scaling regime remains relevant until its intersection with the friction law \eqref{eq:Cf_Prandtl}.
The increase of $C_f$ with $Gr$ in the laminar regime would therefore delay the transition to this `fully turbulent' Prandtl friction law \eqref{eq:Cf_Prandtl} to higher $\Rey$.
At low $\Rey$, although the relationship exhibits a laminar-like scaling, one should recall that the convection flow in the interior remains turbulent.
In figure \ref{fig:Cf_Re_horiz}b, we focus on clarifying this low $\Rey$ regime and the increase in $C_f$ with stronger convection.
Across both mixed convection systems and for a range of $\Pran$, we find a collapse of the data upon rescaling by $Gr^{1/4}$.
The $\Rey^{-1}$ scaling that arises from laminar profiles in Couette/Poiseuille flow, appears somewhat too steep to accurately describe the data.
\citet{blass_flow_2020} reported a scaling of $C_f\sim Re^{-0.90}$ in Couette--Rayleigh--Bénard, but at this time there is no theoretical basis for such a scaling.
Note that one could equivalently express the simplified $\Rey^{-1}$ collapse as $C_f \sim Ri^{1/4}\Rey^{-1/2}$ using the definitions of \eqref{eq:GrRi_def}.
In the case of mixed convection, the buoyancy-driven flow generates non-zero Reynolds stresses in the equation for the mean profile, which can be expected to modify the mean momentum budget and lead to such an increase in $C_f$.
This shall be analysed in further detail for mixed VC in \S\ref{sec:profiles}.

We now turn to the friction coefficient associated with the buoyancy-driven component of the flow.
For the mixed VC system, we can simply take the peak velocity $W_\mathrm{max}$ of the mean vertical velocity $\overline{w}(y)$ as the relevant velocity scale and directly measure the mean vertical shear stress at the wall $\tau = \rho \nu \pd_y \overline{w}$.
In the Rayleigh--B\'enard configuration, the convection has no preferential direction along the walls, resulting in zero mean shear stress.
However, we can still construct a friction coefficient associated with the persistent large-scale circulation by using the root-mean-squared (RMS) horizontal velocity profile $u_H(y) = (\overline{u^2 + v^2})^{1/2}$.
When a horizontal crossflow is added to the RB system, the large-scale circulation aligns itself perpendicular to the imposed flow \citep{pirozzoli_mixed_2017,yerragolam_how_2022}, so for the mixed RB cases we construct a friction coefficient using only the spanwise RMS velocity.
Defining the friction coefficient in this way will only be appropriate for cases where the convectively-driven flow is stronger than the spanwise velocity fluctuations induced by the turbulent shear flow, that is for $Ri\geq O(1)$.

\begin{figure}
    \centering
    \includegraphics{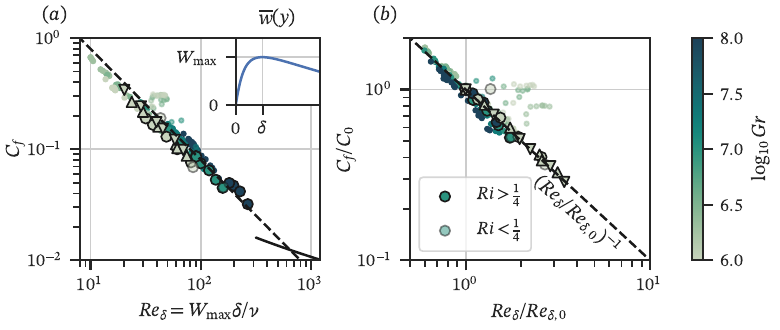}
    \caption{
        $(a)$ Friction coefficients associated with the convectively-driven flow as a function of the boundary layer Reynolds number $Re_\delta=W_\mathrm{max} \delta/\nu$, where $W_\mathrm{max}$ and $\delta$ are computed as in the mean velocity profile shown in the inset.
        $(b)$ The same data normalised against the values from the corresponding natural convection system.
        For the mixed VC cases, different symbols denote different $\Pran$ as outlined in the legend of figure \ref{fig:Cf_Re_horiz}.
        The colour of the markers represents $\Gr$, and simulations with $Ri<1/4$ are plotted as semi-transparent.
        This figure is also available as an \href{https://cocalc.com/share/public_paths/c690a4a6317a519769e8c2b6c3e5c97a40c2338f/figure\%205/figure\%205.ipynb}{interactive JFM notebook}.
    }
    \label{fig:Cf_Re_conv}
\end{figure}

In terms of the Reynolds number, the plate separation $H$ is no longer the appropriate length scale for describing the boundary layer dynamics of the convectively-driven flow.
As shown in the inset of figure \ref{fig:Cf_Re_conv}a, the mean profile $\overline{w}(y)$ of the vertical velocity in (mixed) VC reaches its peak value at a certain wall-normal distance $\delta$.
From this, we can define a boundary layer Reynolds number
\begin{equation}
    Re_\delta = \frac{W_\mathrm{max} \delta}{\nu}, \label{eq:Re_delta_def}
\end{equation}
that drives the behaviour at the wall.
We construct an analogous Reynolds number for the mixed Rayleigh--Bénard system using the spanwise RMS velocity profile.
The friction coefficients of the convective flow component are plotted against $Re_\delta$ in figure \ref{fig:Cf_Re_conv}.
Note that $Re_\delta$ is not known \textit{a priori} but is itself a response parameter of the system that varies with $\Gr$, $\Pran$, and $\Rey$.
Similar to the low $\Rey$ regime for the streamwise friction, we observe a power-law scaling close to $C_f\sim Re_\delta^{-1}$.

This is made clearer in figure \ref{fig:Cf_Re_conv}b, where we collapse the data using the friction coefficient $C_0$ and Reynolds number $Re_{\delta,0}$ obtained from the corresponding natural convection flows, matching $Ra$ and $\Pran$.
Deviations from the scaling relation are observed for cases where $Ri<1/4$, highlighted by translucent symbols in the figure.
As mentioned above, for the mixed RB system this is likely an artefact of using the spanwise RMS velocity to construct the friction coefficient.
However, we also observe the discrepancy for low $Ri$ in mixed VC, suggesting that at low Richardson numbers the turbulence generated by the imposed horizontal flow disrupts the near-wall vertical velocity.
Within the range of parameters explored here, the pressure-driven horizontal flow does not modify the vertical $Re_\delta$ by more than a factor of 4, suggesting that even in the case of mixed convection, the primary control parameters determining $Re_\delta$ are $Gr$ and $Pr$.
As discussed in the appendix of \citet{howland_boundary_2022}, a `fully turbulent' transition of this boundary layer may be possible, analogous to the so-called `ultimate regime' in Rayleigh--Bénard convection \citep{lohse_ultimate_2023}, but only at very high $Gr$.
A more in-depth analysis of the mean vertical momentum budget in mixed VC will be presented in \S\ref{sec:profiles}.

\subsection{Nusselt number}
The dimensionless heat flux through the system is characterised by the Nusselt number, defined as
\begin{align}
    \Nu &= \frac{F_\theta}{\kappa \Delta / H}, &
    F_\theta &= -\kappa \frac{\pd \overline{\theta}}{\pd y} + \overline{v'\theta'} .
\end{align}
Here $F_\theta$ is the horizontal heat flux through the system (normalised by the specific heat $\rho c_p$), and the overbar denotes averaging in time and in directions parallel to the plates.
Integration of the mean temperature equation \eqref{eq:heat_flux_def} shows that $F_\theta$ is constant across the domain in a statistically steady state.

\begin{figure}
    \centering
    \includegraphics{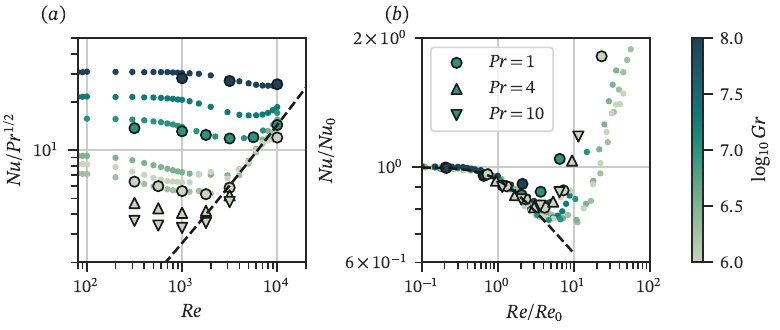}
    \caption{
        Nusselt numbers plotted as a function of Reynolds number.
        $(a)$ The Nusselt number is normalised by $\Pran^{1/2}$ to focus on the response in the shear-dominated regime at high $\Rey$. The black dashed line marks the Reynolds analogy $Nu\approx \frac{1}{4}C_f \Rey \Pran^{1/2}$, where $C_f$ satisfies the Prandtl friction law \eqref{eq:Cf_Prandtl}.
        $(b)$ The data is normalised by the values associated with natural convection.
        Here, the black dashed line marks the recently proposed scaling relation $\Nu/\Nu_0 \sim (\sqrt{1 + (\Rey/\Rey_0)^2})^{-1/5}$ from \citet{yerragolam_scaling_2024}.
        This figure is also available as an \href{https://cocalc.com/share/public_paths/c690a4a6317a519769e8c2b6c3e5c97a40c2338f/figure\%206/figure\%206.ipynb}{interactive JFM notebook}.
    }
    \label{fig:Nuss_comparison}
\end{figure}

In figure \ref{fig:Nuss_comparison}$(a)$, we plot the Nusselt number compensated by $\Pran^{1/2}$.
This pre-factor seems appropriate for the $\Pran$-dependence of passive scalar transport in turbulent boundary layers when $\Pran\lesssim O(1)$ \citep{kays_convective_2005}, although at higher $\Pran$ one expects a transition towards a $\Pran^{1/3}$ dependence \citep{kader_heat_1972,alcantara-avila_direct_2021}.
The data of both systems converge towards the expression
\begin{equation}
    Nu\approx \frac{C_f(\Rey)}{4} \Rey \Pran^{1/2} , \label{eq:Nu_Prandtl}
\end{equation}
where $C_f(\Rey)$ follows the Prandtl law of \eqref{eq:Cf_Prandtl}.
This expression draws a parallel between the transport of heat and momentum at the wall, known as the Reynolds analogy, and describes the heat transport in `forced convection' when buoyancy no longer affects the flow.
From this data, we anticipate that the forced convection expression \eqref{eq:Nu_Prandtl} applies for Reynolds numbers greater than that where the friction coefficient begins to follow the turbulent friction law shown in figure \ref{fig:Cf_Re_horiz}.
At low $\Rey$ (or more precisely high $Ri$), the Nusselt number responses of the two systems (VC and RB) do not match as precisely as the friction coefficients.
Indeed, in the absence of an external flow, VC and RB do not exhibit the same $Nu(Ra,\Pran)$ response due to the lack of coupling between the kinetic energy budget and the heat flux in VC \citep{ng_vertical_2015}.

However, we observe a more universal behaviour when the Nusselt numbers in the mixed convection systems are normalised by the values for the equivalent natural convection systems $\Nu_0(\Gr, \Pran, \Rey) = \Nu(\Gr, \Pran)|_{\Rey=0}$.
In figure \ref{fig:Nuss_comparison}$(b)$, the normalised Nusselt numbers are plotted as a function of the input Reynolds number normalised by the Reynolds number of the natural convection case $\Rey_0=W_0H/\nu$.
For the mixed VC cases, we define $W_0(\Gr, \Pran, \Rey) = W_\mathrm{max}(\Gr, \Pran)|_{\Rey=0}$ as the peak velocity of the natural VC flow as highlighted in the inset of figure \ref{fig:Cf_Re_conv}$(a)$.
For mixed RB, we follow \citet{yerragolam_scaling_2024} in using the volume-averaged RMS velocity for $W_0$, and note that the results are insensitive to this choice of velocity scale in describing the `wind' of the large-scale circulation.
For $\Rey/Re_0=O(1)$, all the data from both configurations collapse onto a single curve, showing a drop in the heat flux of up to 25\%.
Given this collapse, the $Re_0$ of the natural convection appears to be a critical $Re$ above which the Nusselt number is significantly affected by the horizontal crossflow.
\citet{yerragolam_scaling_2024} provide an estimate for the drop in $Nu$, derived from the kinetic energy balance in mixed Rayleigh--Bénard, but this balance cannot be related to the \emph{horizontal} heat flux that is relevant for mixed VC.

In summary, in this section we have demonstrated the universality in the global response parameters of mixed RB and mixed VC, namely in the friction coefficient $C_f$ and the Nusselt number, and the limitations of this universality.
In the following sections, we will compare more local quantities, starting with the wall-normal profiles.

\section{Wall-normal profiles in mixed vertical convection \label{sec:profiles}}

We now turn to the first and second order statistics, averaged parallel to the plates, to further investigate the dynamics behind the observed global responses.
For clarity, we focus solely on the new simulations of mixed vertical convection and study the variation across the three-parameter space of $\Gr$, $\Pran$ and $\Rey$.

\begin{figure}
    \centering
    \includegraphics{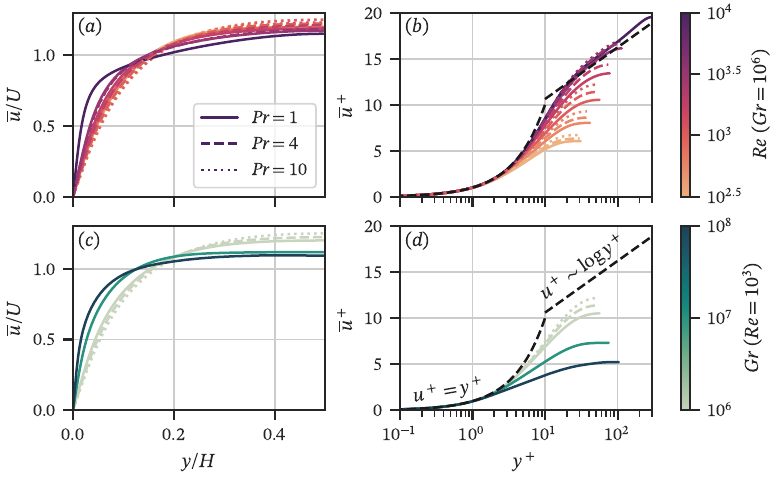}
    \caption{
        Mean profiles of the streamwise velocity $\overline{u}(y)$.
        $(a,b)$ present data from simulations at a fixed $Gr=10^6$, varying $\Rey$ and $\Pran$. $\Rey$ variation is denoted by the line colour and $\Pran$ variation is denoted by the line style.
        $(c,d)$ present data from simulations at a fixed $\Rey=10^3$, varying $\Gr$ and $\Pran$.
        In the left column, the profiles are normalised by the imposed bulk velocity $\langle u\rangle = U$ and the plate separation $H$.
        In the right column, profiles are presented in viscous wall units, where $\overline{u}^+ = \overline{u}/{u_\tau}$ and $y^+=y u_\tau/\nu$.
        Dashed black lines represent the linear relation $u^+=y^+$ and the logarithmic region $u^+ = \kappa_u^{-1} \log{y^+} + B$ where the von Kármán constant $\kappa_u = 0.41$ and $B=5$ are taken from \citet{pirozzoli_turbulence_2014}.
        This figure is also available as an \href{https://cocalc.com/share/public_paths/c690a4a6317a519769e8c2b6c3e5c97a40c2338f/figure\%207/figure\%207.ipynb}{interactive JFM notebook}.
    }
    \label{fig:u_profiles}
\end{figure}

We begin with the response of the mean streamwise velocity $\overline{u}(y)$ in figure \ref{fig:u_profiles}.
For a fixed $Gr=10^6$, as in panels $(a)$-$(b)$, the effect of increasing $\Rey$ can be seen most clearly when the mean velocity profiles are scaled by viscous wall units in figure \ref{fig:u_profiles}$(b)$.
As $\Rey$ increases, the velocity profile tends towards the classical log-law profile, with the case of $\Rey=10^4$ closely matching the profile of turbulent Poiseuille flow, as in for example \citet{lee_direct_2015}.
The effect of stronger thermal convection on the mean profile is also similar to that observed in other mixed convection systems in the literature \citep{scagliarini_law_2015,blass_flow_2020}.
In figure \ref{fig:u_profiles}$(c)$, where $\Rey=10^3$ is fixed and $\Gr$ varies between $10^6$ and $10^8$, higher $\Gr$ leads to a flatter mean profile in the bulk of the channel.
This is illustrated further in wall units in figure \ref{fig:u_profiles}$(d)$, where a significant drop in $\overline{u}^+$ is observed for $y^+ =O(10)$.
Plus symbols denote scaling in viscous wall units with velocity $u_\tau = \sqrt{\tau_u/\rho}$ and length $\nu/u_\tau$.
Such a drop is consistent with the previous findings of \citet{scagliarini_law_2015} for mixed Rayleigh--Bénard, who proposed a modified log-law based on mixing length arguments coupled to the temperature field.

\begin{figure}
    \centering
    \includegraphics{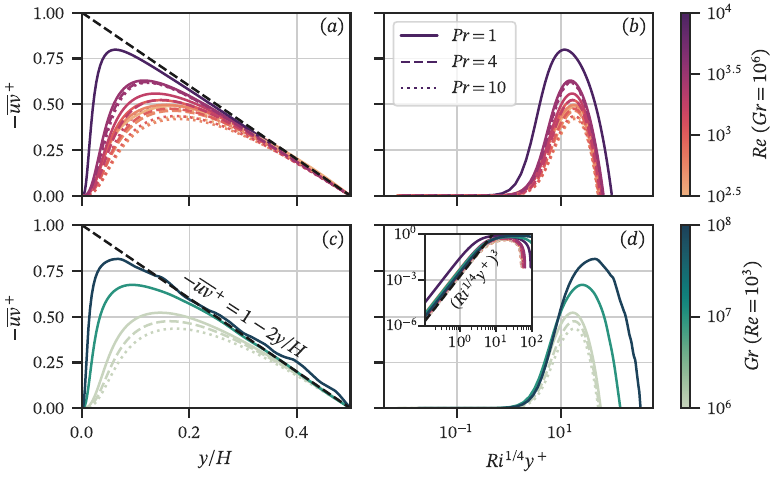}
    \caption{
        Wall-normal profiles of the streamwise Reynolds stress component $\overline{uv}(y)$ scaled by the streamwise friction velocity $\overline{uv}^+ = \overline{uv}/u_\tau^2$.
        As in figure \ref{fig:u_profiles}, panels $(a,b)$ are at fixed $Gr=10^6$ and panels $(c,d)$ are at fixed $\Rey=10^3$.
        Colours and line styles are as detailed in the caption of figure \ref{fig:u_profiles}.
        The left column presents profiles relative to the plate separation $H$, whereas the right column shows the profiles in viscous wall units scaled by $Ri^{1/4}$.
        The inset of $(d)$ presents the data of panels $(b,d)$ with a logarithmic scale on both axes to highlight the near-wall collapse of the data.
        This figure is also available as an \href{https://cocalc.com/share/public_paths/c690a4a6317a519769e8c2b6c3e5c97a40c2338f/figure\%208/figure\%208.ipynb}{interactive JFM notebook}.
    }
    \label{fig:uv_profiles}
\end{figure}

Further insight for the streamwise velocity can be gained from the appropriate component of the Reynolds stress.
Considering a statistically steady state, we average the streamwise component of the momentum equation \eqref{eq:NSmom} to obtain
\begin{equation}
    \pd_y \overline{uv} = \nu \pd_{yy} \overline{u} + \overline{G}, \label{eq:mean_mom}
\end{equation}
where an overbar denotes an average in the periodic ($x$, $z$) directions and in time.
Note that wall-normal advective fluxes in incompressible channel flows are purely turbulent, i.e.~$\overline{uv}\equiv \overline{u'v'}$ since $\overline{v}\equiv 0$.
From volume-averaging, we can also relate the mean pressure gradient forcing to the mean wall shear stress through $\overline{G}=2\tau_u/\rho H$.
The first integral of \eqref{eq:mean_mom} can therefore be written as
\begin{equation}
    \pd_{y^+} \ubar^+ - \overline{uv}^+ =  1 - \frac{2y}{H} . \label{eq:mean_mom_int}
\end{equation}
From \eqref{eq:mean_mom}-\eqref{eq:mean_mom_int}, the close coupling of the mean streamwise velocity and the Reynolds stress $\overline{uv}$ is evident.
We therefore present the profiles of $\overline{uv}(y)$ in figure \ref{fig:uv_profiles}.
As expected, the Reynolds stress dominates the viscous contribution to \eqref{eq:mean_mom_int} away from the walls, leading to a balance of $-\overline{uv}^+\approx 1 - 2y/H$ as shown in panels $(a)$ and $(c)$.
Relative to $H$, the boundary layer in which the viscous term is relevant becomes thinner as both $\Rey$ and $\Gr$ increase.
The near-wall behaviour of $\overline{uv}$ exhibits a remarkable collapse when scaled by $Ri^{1/4}$ as in panels $(b)$ and $(d)$ of figure \ref{fig:uv_profiles}, except for the highest $Re$ case with $Ri=0.01$.

The additional factor $Ri^{1/4}$ suggests that the appropriate near-wall length scale for the Reynolds stress is modified from the standard viscous wall unit as
\begin{align}
    Ri^{1/4} y^+ &= \frac{y}{l}, &
    l &= \frac{\nu}{\sqrt{\frac{U_f}{U}\frac{\tau_u}{\rho}}}.
\end{align}
The additional pre-factor of $U_f/U$ in front of the shear stress suggests that the vertical, convectively-driven component of shear cannot be neglected when considering the streamwise Reynolds stress.
An improved collapse to that seen in figure \ref{fig:uv_profiles}$(c,d)$ can be found by computing a viscous length scale using the total shear stress at the wall $\tau = \sqrt{\tau_u^2 + \tau_w^2}$.
As explicitly shown in appendix \ref{app:wall_scaling}, with this scaling the Reynolds stress for the highest $Re$ case also matches the other curves.
The presence of the convectively-driven flow increases the near-wall Reynolds stress, which in turn leads to the flattened mean velocity profiles observed in figure \ref{fig:u_profiles}d.
This result qualitatively explains the origin of the change in $C_f$ with $Gr$ seen in figure \ref{fig:Cf_Re_horiz}, where enhanced buoyancy driving led to a larger skin friction.
The increase in near-wall Reynolds stress that arises due to convection thins the boundary layer of the mean horizontal velocity, which in turn produces a larger mean gradient at the wall and a larger friction coefficient $C_f$.

\begin{figure}
    \centering
    \includegraphics{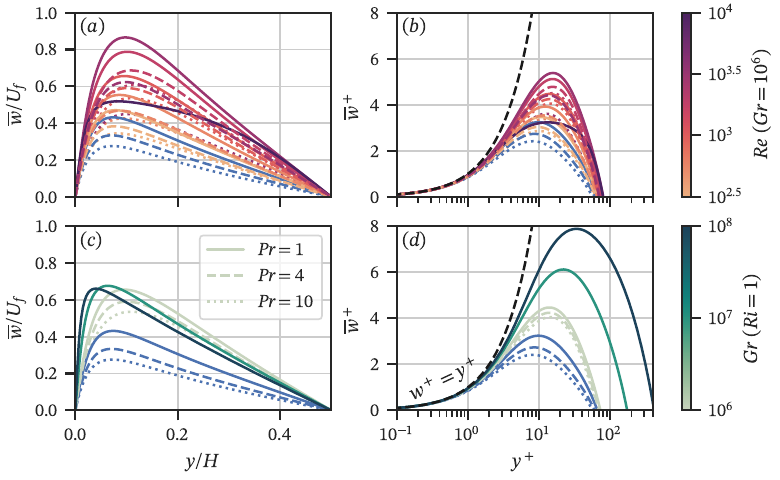}
    \caption{
        Mean profiles of the vertical velocity $\overline{w}(y)$.
        As in figure \ref{fig:u_profiles}, panels $(a,b)$ are at fixed $\Gr=10^6$, but here panels $(c,d)$ are at fixed $Ri=1$.
        In the left column, the profiles are normalised by the free-fall velocity $U_f = \sqrt{g\alpha \Delta H}$ and the plate separation $H$.
        In the right column, profiles are presented in viscous wall units, where $\overline{w}^+ = \overline{w}/{w_\tau}$ and $y^+=y w_\tau/\nu$.
        Reference data for natural vertical convection (with $Gr=10^6$, $\Rey=0$) are shown in blue.
        This figure is also available as an \href{https://cocalc.com/share/public_paths/c690a4a6317a519769e8c2b6c3e5c97a40c2338f/figure\%209/figure\%209.ipynb}{interactive JFM notebook}.
    }
    \label{fig:w_profiles}
\end{figure}

\begin{figure}
    \centering
    \includegraphics{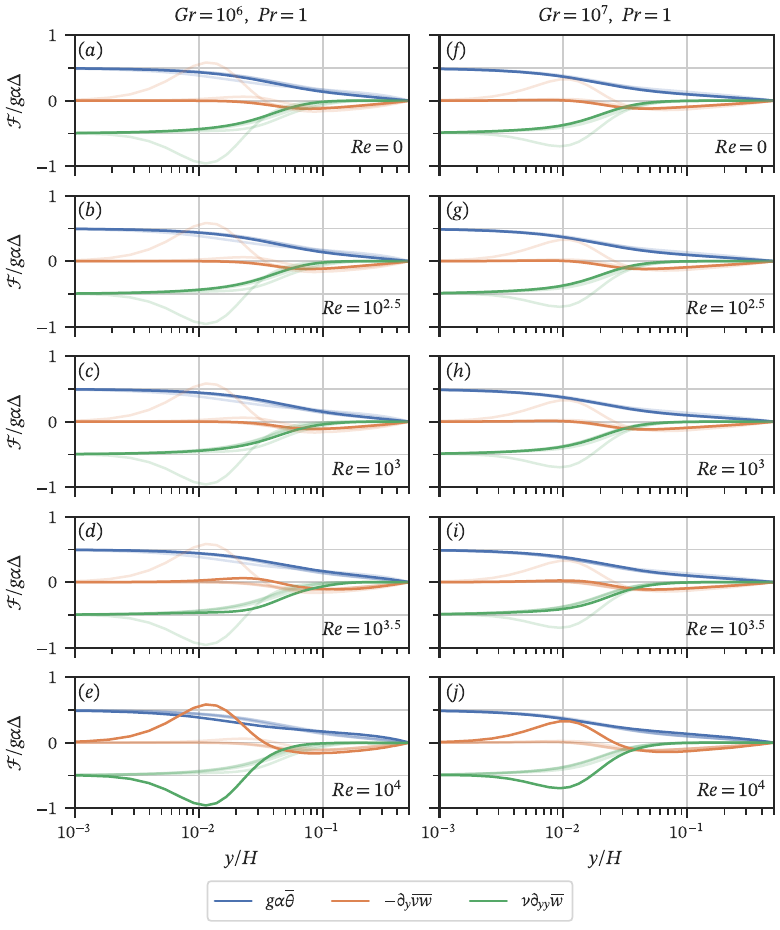}
    \caption{
        Profiles of the mean vertical momentum budget terms for $\Pran=1$ and $Gr=10^6$ (left column) or $Gr=10^7$ (right column). Reynolds numbers vary from $(a,f)$ $\Rey=0$ to $(e,j)$ $\Rey=10^4$.
        Colours denote the budget term being plotted as detailed in the legend, and each budget term is plotted normalised against the buoyancy scale $g\alpha \Delta$.
        Within each column, semi-transparent lines are added to show the profiles for the Reynolds numbers highlighted by the other rows as comparison.
        The wall-normal coordinate $y$ is plotted on a logarithmic axes to highlight the variation in the near-wall region.
        This figure is also available as an \href{https://cocalc.com/share/public_paths/c690a4a6317a519769e8c2b6c3e5c97a40c2338f/figure\%2010/figure\%2010.ipynb}{interactive JFM notebook}.
    }
    \label{fig:vmom_budget}
\end{figure}

The effect of convection on the shear is by no means a one-way interaction.
This is evident from the modification of the mean vertical velocity profile by the imposed horizontal flow.
In figures \ref{fig:w_profiles}$(a,b)$, vertical velocity profiles are shown for fixed $\Gr=10^6$ and varying $\Rey$, with the reference natural convection case ($\Rey=0$) highlighted in blue for comparison.
Compared to the natural VC case, the introduction of horizontal driving at moderate $\Rey$ leads to an increase in the peak vertical velocity, and hence an increase in the mean shear both in the bulk and at the walls.
At the highest $\Rey=10^4$ (the darkest red line in figure \ref{fig:w_profiles}), a subsequent decrease is observed in the peak vertical velocity, as well as a nonlinear profile in the bulk.
None of the cases studied here exhibit a log-layer in the vertical velocity, with the largest $\overline{w}^+$ being approximately 6.5 for the most strongly convective case of $Gr=10^8$.
For fixed $Ri=1$ (shown in figures \ref{fig:w_profiles}$(c,d)$), all cases show a similar increase in the peak velocity, and the mean gradient in the bulk appears largely independent of $\Gr$ and $\Pran$.
The distance of the velocity peak from the wall (compared to the channel width $H$) decreases for larger $\Gr$, but the value of the peak velocity in free-fall units does not strongly depend on $\Gr$.
This similarity at constant $Ri$ is suggestive that the vertical velocity modification is primarily determined by the \emph{relative} strength of the horizontal flow to convection.

\begin{figure}
    \centering
    \includegraphics{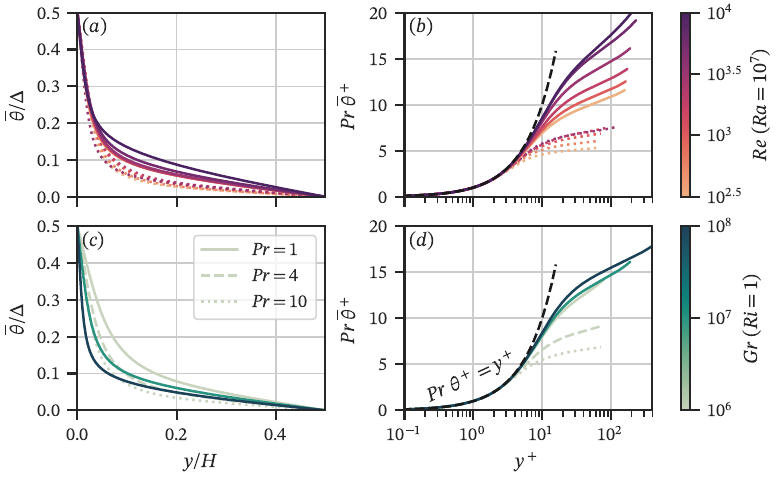}
    \caption{
        Mean temperature profiles for $(a,b)$ fixed $Ra=10^7$ and $(c,d)$ fixed $Ri=1$.
        As in figures \ref{fig:u_profiles}-\ref{fig:w_profiles}, colours denote changes in $\Rey$ and $\Gr$, whereas line styles show changes in $\Pran$.
        In the left column, temperature is normalised by the difference across the plates $\Delta$ and on the right, temperature is shown in terms of wall units $\theta^+=(\theta_w - \theta) U_\tau / F_\theta$, where $\theta_w$ is the value of temperature at the wall $y=0$.
        The $U_\tau$ used here and in $y^+=yU_\tau/\nu$ is calculated using both components of the wall shear stress, as defined in \eqref{eq:total_Utau}.
        This figure is also available as an \href{https://cocalc.com/share/public_paths/c690a4a6317a519769e8c2b6c3e5c97a40c2338f/figure\%2011/figure\%2011.ipynb}{interactive JFM notebook}.
    }
    \label{fig:T_profiles}
\end{figure}

To further investigate the behaviour of the vertical velocity profiles, we now turn to the mean vertical momentum budget.
Unlike the streamwise velocity in \eqref{eq:mean_mom}-\eqref{eq:mean_mom_int}, the vertical velocity is not only tied to the Reynolds stress profile.
Rather, the mean vertical momentum equation reads
\begin{equation}
    \pd_y \overline{vw} = \nu \pd_{yy} \wbar + g\alpha \overline{\theta} \label{eq:mean_vmom} .
\end{equation}
Close to the wall, we expect the Reynolds stress to be negligible and a balance to arise between buoyancy and viscosity.
Due to the symmetry of the boundary conditions, all three terms must be zero at the channel centre.
In the bulk of the flow, the mean velocity is approximately linear, so we expect a balance between buoyancy and Reynolds stress.
These features are present in each of the simulations highlighted in figure \ref{fig:vmom_budget}.
As $\Rey$ increases, the key modification to the budget arises in the Reynolds stress term $-\pd_y \overline{vw}$.
In natural VC, there is a miniscule positive peak in the Reynolds stress term close to the wall, but its amplitude is so small that it is indistinguishable in figures \ref{fig:vmom_budget}$(a,f)$.
This peak grows with $\Rey$, becoming visible at $\Rey=10^{3.5}$ in panels $(d,i)$, coinciding with a flattened profile of the viscous term (shown in green).
By $\Rey=10^4$, at which the Reynolds stresses are significantly energised by the horizontal forcing, the nonlinear term peak exceeds the contribution from the buoyancy term.
This leads to a significant drop in the viscous term around $y\approx 10^{-2}H$, which determines the modified mean velocity observed in figure \ref{fig:w_profiles}$(a,b)$.

Changes in the mean temperature profile $\overline{\theta}$ in figure \ref{fig:vmom_budget} are more subtle, with the most obvious feature being a slight drop in panel $(e)$, coinciding with the Reynolds stress peak.
A clearer picture of the mean temperature response can be found in figure \ref{fig:T_profiles}, where profiles are plotted both on linear axes and in wall units.
Since the dimensionless wall temperature gradient is equivalent to the Nusselt number, which is most strongly dependent on $Ra$, we compare the temperature profiles in panels $(a,b)$ at varying $\Rey$ and $\Pran$ but fixed $Ra=10^7$.
As shown in these panels, the temperature gradient in the bulk increases with $\Rey$, and any possible log-law profile does not collapse to a universal slope or coefficient.
For fixed $Ri=1$ however, shown in figure \ref{fig:T_profiles}$(c,d)$, the bulk gradient appears independent of $\Gr$, suggesting that $Ri$ and $\Pran$ are the key control parameters determining the flow properties away from the walls.

The mean profiles presented here do not further clarify the origin of the non-monotonic behaviour of $Nu$ with $Re$ seen in figure \ref{fig:Nuss_comparison}.
When normalised by the wall heat flux, the mean temperature profiles in figure \ref{fig:T_profiles}(b) show a monotonic increasing trend with $Re$ away from the wall.
The non-monotonic behaviour observed in the vertical velocity profiles of figure \ref{fig:w_profiles}(a,b) suggests an intrinsic connection between the vertical flow and the wall-normal heat transport, although the very subtle changes in the vertical momentum budget remain difficult to interpret in the context of the non-monotonic Nusselt number variation.
The later sections on spectral analysis and statistics of the boundary layer aim to shed more light on this global heat transport.

To conclude our analysis of the mean profiles, we compare the new simulations of mixed vertical convection to the universal functions of the Monin--Obukhov similarity theory \citep{monin_basic_1954,foken_50_2006}.
As outlined in the introduction, the theory considers a region of the flow where viscosity and molecular diffusion are negligible and the appropriate length scale is the Obukhov length
\begin{equation}
    L_O=\frac{u_\tau^3}{g\alpha F_\theta} \label{eq:obukhov_length} .
\end{equation}
On dimensional grounds, universal functions that depend only on $y/L_O$ for the first and second order statistics can then be constructed for the temperature and velocity fields.
One key assumption used to derive the universal functions for velocity is that the turbulent momentum flux $\rho \overline{uv}$ is constant over the region of interest.
However, as shown in figure \ref{fig:uv_profiles}, this is not true in a turbulent channel flow driven by a pressure gradient, where the Reynolds stress follows a linear relation in the bulk.
We can therefore only expect the universal functions of \citet{monin_basic_1954} to hold for a small range of $y$ in our simulations.

\begin{figure}
    \centering
    \includegraphics{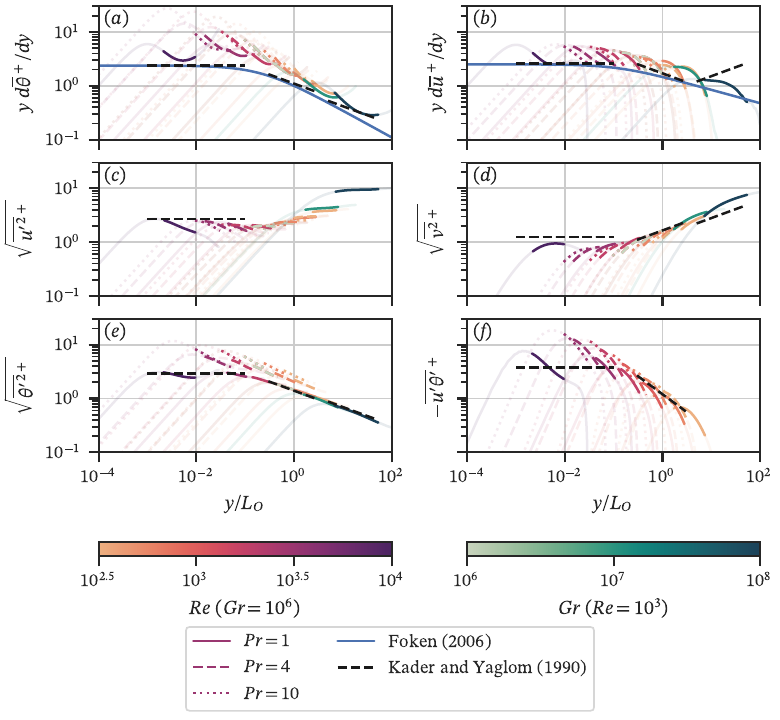}
    \caption{
        Comparison of the mixed VC simulation results with the `universal functions' of the Monin--Obukhov similarity theory for first and second order statistics.
        Profiles are shown as semi-transparent unless the advective wall-normal fluxes of heat $\overline{v \theta}$ and momentum $\overline{v u}$ are within 80\% of their maximum value.
        Consistent with the Monin--Obukhov formulation, all quantities here are normalised by the horizontal component of the friction velocity $u_\tau$ and the wall-normal heat flux $F_\theta$.
    }
    \label{fig:monin-obukhov}
\end{figure}

Considering these caveats, we present a collection of mean profiles from the mixed vertical convection simulations as a function of $y/L_O$ in figure \ref{fig:monin-obukhov}.
Following \citet{pirozzoli_mixed_2017}, who present a similar figure (their figure 16) for mixed Rayleigh--Bénard convection, we only emphasise the region of the domain where the turbulent fluxes of heat and momentum are greater than 80\% of their maximum value, with the rest of the profiles made semi-transparent.
We compare our results with the updated scaling theories of \citet{kader_mean_1990} and, for the mean temperature and velocity profiles, the commonly used Businger--Dyer relations outlined in \citet{foken_50_2006}.
For the highlighted region where the fluxes are close to their peak, the majority of the results collapse onto a single function of $y/L_O$ close to these profiles.
In figure \ref{fig:monin-obukhov}(a), there are two notable discrepancies from the theoretical estimates for the mean temperature profile.
Firstly, the values at transitional $y/L_O\approx 10^{-1}$ are roughly a factor 2 larger than the suggested profiles.
Secondly, the effective power law scaling observed for strong buoyancy driving $y\geq L_O$ appears steeper than the $-1/3$ proposed by \citet{kader_mean_1990} (dashed black lines), and closer to the $-1/2$ scaling associated with the asymptotic limit of the Businger--Dyer relation (solid blue lines).
A discrepancy in the two theoretical predictions for the mean velocity profile at $y\gg L_O$, highlighted in figure \ref{fig:monin-obukhov}(b), cannot be resolved from the results presented here, although we note that the data here agrees well with the mixed RB data of \citet{pirozzoli_mixed_2017}.
The variation in $Pr$ across the simulations collapses for most quantities in figure \ref{fig:monin-obukhov}, except for the rms temperature shown in panel (e).
The origin of this discrepancy with $Pr$ is currently unclear, but understanding how the $Pr$-dependence may affect the Monin--Obukhov profiles is important in the context of their applications outside the atmosphere, for example in the surface layer of the upper ocean \citep{zheng_evaluating_2021}.

\section{Spectral analysis \label{sec:spectra}}

We now investigate the scale-dependence of the thermal structures and heat flux in mixed convection through analysis of the power spectra.
To ensure that we capture the full range of dynamical scales, we perform further simulations in extended aspect ratio domains, with $L_x=L_z=24 H$.
The details of these simulations are outlined in table \ref{tab:spec_params} of appendix \ref{app:numerics}.

In unsheared Rayleigh--Bénard convection, \citet{krug_coherence_2020} show that large-scale patterns, also known as superstructures, can be identified from a low-wavenumber peak in the power spectrum of the temperature field and in the co-spectrum of the wall-normal heat flux.
This peak denotes the scale of the large-scale circulation or `wind' of convection key to theories describing RBC.
Since there is no preferential horizontal flow direction in RBC, one-dimensional spectra can be analysed, but in mixed and vertical convection, the two directions parallel to the plates must be considered separately.

We therefore compute time-averaged spectra using two-dimensional Fourier transforms in the periodic directions.
For visualisation purposes, we present separate one-dimensional spectra for the streamwise ($x$) and spanwise ($z$) wavenumbers, which are computed by integrating over the other wavenumber.
Precisely, the two- and one-dimensional co-spectra of any two variables $f$ and $g$ are defined as
\begin{align}
    \Phi_{fg}(k_x, y, k_z) &= \mathbb{R}\left[\hat{f}^* \hat{g}\right], &
    \Phi_{fg}(k_x,y) &= \int \mathbb{R}\left[\hat{f}^* \hat{g}\right] \, \mathrm{d}k_z,
\end{align}
where $\hat{f}$ denotes the two-dimensional Fourier transform of $f$ in $x$ and $z$, $\mathbb{R}$ is the real part.
The Fourier transforms are normalised such that integrating the spectrum over wavenumber space recovers the corresponding volume-averaged quantity $\langle f g \rangle = \iint \Phi_{fg} \mathrm{d}k_x \mathrm{d}k_z$.

\begin{figure}
    \centering
    \includegraphics{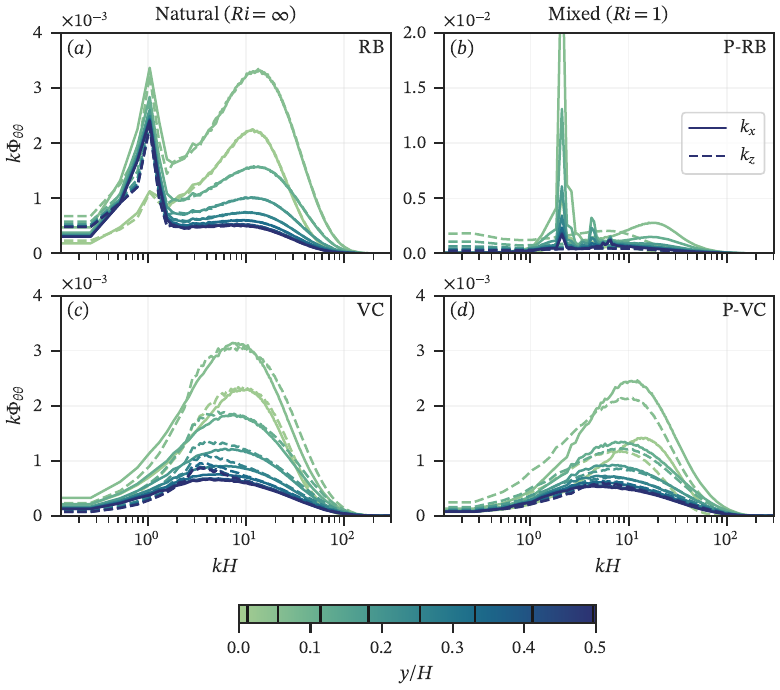}
    \caption{
        One-dimensional power spectra $\Phi_{\theta \theta}(k,y)$ of temperature for the extended domain simulations described in table \ref{tab:spec_params}.
        Solid lines denote spectra as a function of $k=k_x$ and dashed lines are functions of $k=k_z$.
        The colour of each line is determined by its wall normal location $y$, with the specific locations highlighted on the colour bar.
        Each spectrum is multiplied by the wavenumber $k$ such that area under each curve is representative of the relative contribution in wavenumber space.
        The left column considers simulations of natural convection ($\Rey=0$), whereas the right column presents simulations of mixed convection at $Ri=1$.
        The upper row shows RB cases with gravity in the wall normal ($y$) direction, and the bottom row show VC cases with gravity parallel to the wall in $z$.
        All simulations have fixed $Gr=10^7$, $\Pran=1$.
        This figure is also available as an \href{https://cocalc.com/share/public_paths/c690a4a6317a519769e8c2b6c3e5c97a40c2338f/figure\%2012/figure\%2012.ipynb}{interactive JFM notebook}.
    }
    \label{fig:temp_spectra}
\end{figure}

The power spectrum of temperature $\Phi_{\theta \theta}$ is presented in figure \ref{fig:temp_spectra} for the four extended simulations.
In panel $(a)$, for the standard RB configuration, we see similar behaviour to \citet{krug_coherence_2020}, with a distinct, sharp low wavenumber peak at $k H\approx 1$, and a more broad peak at smaller scales.
As mentioned above, the lack of a preferential direction means that the two directional spectra are virtually identical.
When shear is added to the RB system in panel $(b)$, the horizontal isotropy in the system is destroyed.
As expected from previous work on mixed RB \citep{pirozzoli_mixed_2017,blass_flow_2020,blass_effect_2021,yerragolam_how_2022}, coherent rolls aligned with the streamwise axis dominate the signal.
Note the difference in $y$-axis between figures \ref{fig:temp_spectra}$(a)$ and \ref{fig:temp_spectra}$(b)$.
In the streamwise spectrum $\Phi_{\theta\theta}(k_z,y)$ a sharp peak at $kH\approx 2$ is visible at all wall-normal locations, but is particularly prominent in the near-wall region (light green) where there is a maximum in $\overline{\theta^2}(y)$.
The corresponding wavelength of these structures is $\lambda = 2\pi H/k_z \approx 3 H$, which is consistent with that observed previously in the mixed convection literature at $Ri=1$ \citep[e.g.][]{pirozzoli_mixed_2017}.
The alignment of the flow structures in the streamwise direction leads to a broad contribution to the streamwise spectrum at low $k_x$.
This behaviour is purely a result of the regular alignment, and is not related to any domain size effects.

In figures \ref{fig:temp_spectra}$(c,d)$, we present the same analysis but for natural and mixed \emph{vertical} convection.
The contrast to the RB cases is immediately apparent, with no sharp peaks at any wavenumber or wall-normal position.
This confirms the earlier visual observation of figure \ref{fig:midplanes}, where the instantaneous snapshots showed no clear coherent length scale.
In natural VC, shown in figure \ref{fig:temp_spectra}$(c)$, the two 1-D spectra do not overlap exactly like in the RB case due to the buoyancy-driven mean flow in the vertical.
The $k_x$ spectrum shows a small peak around $kH\approx 3$ that is absent from the $k_z$ spectrum, and becomes more prominent towards the channel centre.
Comparing figures \ref{fig:temp_spectra}$(c)$ and \ref{fig:temp_spectra}$(d)$, we see that this peak is suppressed by the addition of external shear.
Furthermore, all of the broad mid-range peaks in the spectra exhibit a decrease in amplitude and a shift to higher wavenumbers.

\begin{figure}
    \centering
    \includegraphics{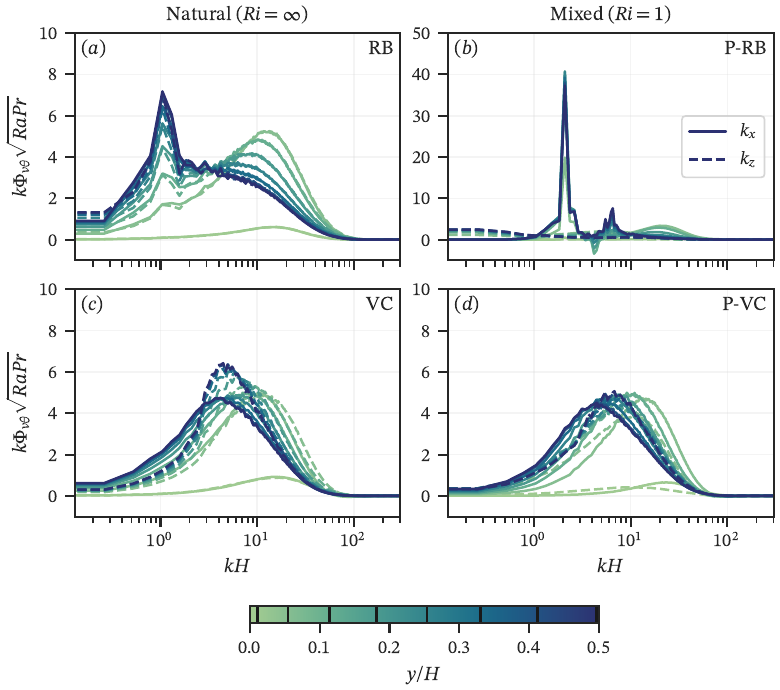}
    \caption{
        One-dimensional co-spectra $\Phi_{v\theta}(k,y)$ of the wall-normal heat flux for the extended domain simulations described in table \ref{tab:spec_params}.
        See the caption of figure \ref{fig:temp_spectra} for more details on the meaning of line styles and colours.
        The spectra are normalised by $\sqrt{RaPr}$ such that integration over $k$ recovers the dimensionless advective heat flux $Nu-1 = \langle v\theta\rangle/(\kappa \Delta /H)$.
        As in figure \ref{fig:temp_spectra}, the configurations presented are $(a)$ RB, $(b)$ mixed RB, $(c)$ VC, $(d)$ mixed VC.
        This figure is also available as an \href{https://cocalc.com/share/public_paths/c690a4a6317a519769e8c2b6c3e5c97a40c2338f/figure\%2013/figure\%2013.ipynb}{interactive JFM notebook}.
    }
    \label{fig:flux_cospectra}
\end{figure}

The heat flux cospectra $\Phi_{v\theta}$, shown in figure \ref{fig:flux_cospectra}, exhibit similar features to the power spectra.
In panels $(a)$ and $(b)$, a sharp low-wavenumber peak is once again visible for RB and mixed RB systems, and for the heat flux this peak increases towards the channel centre.
A broad, high-wavenumber peak is also observed in the spectrum close to the wall.
This peak flattens and shifts to lower wavenumbers (i.e.~larger scales) as distance from the wall increases, which can be interpreted as the emergence and coalescence of small-scale plume structures.
A similar behaviour can be found in the VC configurations of panels $(c)$ and $(d)$, with the broad peak in both $k_x$ and $k_z$ spectra shifting to lower wavenumbers for larger $y$.
However, in the $k_x$ spectrum for natural VC (where $x$ is perpendicular to gravity), this peak also increases in amplitude towards the channel centre, highlighting that the majority of the wall-normal heat transport at the channel centre occurs due to structures of size $\lambda\approx H$.
Comparing this to the mixed VC case in figure \ref{fig:flux_cospectra}$(d)$, the mid-scale peak appears significantly suppressed when the external shear is imposed, suggesting that the coalescence of plumes in the bulk is disrupted by the shear.
This is consistent with the interpretation of \citet{domaradzki_direct_1988} and \citet{scagliarini_heat-flux_2014} who argued that the drop in heat flux in mixed RB can be related phenomenologically to disruption of the organisation of small-scale convective plumes.

\section{Heat flux statistics in the boundary layer \label{sec:BL_stats}}

\begin{figure}
    \centering
    \includegraphics{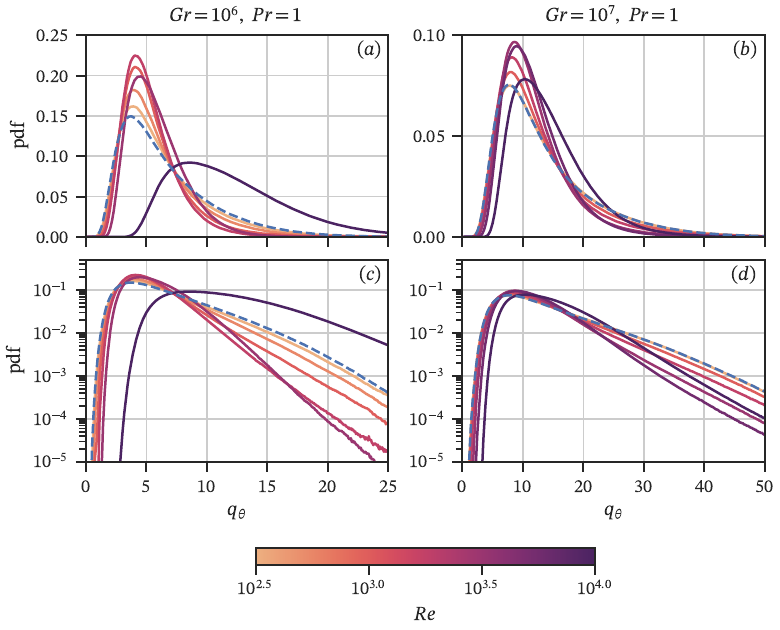}
    \caption{
        Probability density functions of the dimensionless local heat flux $q_\theta$ as defined in \eqref{eq:heat_flux_def} for $Gr=10^6$ (left column) and $Gr=10^7$ (right column).
        Line colours highlight variation in the Reynolds number, as shown by the colour bar, with the blue dashed line plotting the data for natural convection with $\Rey=0$.
        $(a,b)$ are presented with linear $y$-axes, and $(c,d)$ present the same data on logarithmic $y$-axes to highlight the exponential tails.
        This figure is also available as an \href{https://cocalc.com/share/public_paths/c690a4a6317a519769e8c2b6c3e5c97a40c2338f/figure\%2014/figure\%2014.ipynb}{interactive JFM notebook}.
    }
    \label{fig:local_flux_pdf}
\end{figure}

Whereas the previous section focused on the advective heat flux in the bulk and its modification due to shear, we now turn to the conductive heat flux that dominates the transport in the boundary layer.
Specifically, we investigate the statistical distribution of the local conductive heat flux at the boundary plates.
As defined earlier in \eqref{eq:heat_flux_def}, we consider the local dimensionless heat flux $q_\theta$ whose time- and plane-average is equivalent to the Nusselt number.
The pointwise data of $q_\theta(x,z,t)$, also shown in figure \ref{fig:wall_planes}, is collected over time and space to construct histograms that are normalised to produce probability density functions in figure \ref{fig:local_flux_pdf}.
The data shown here is only for $\Pran=1$ and $Gr=10^6, 10^7$ since these cases cover the widest range of Richardson number, highlighting the transition from natural convection to forced convection.

In figure \ref{fig:local_flux_pdf}$(a,b)$, where the data are presented on linear axes, we find the same effect of increasing shear at both Grashof numbers.
As $\Rey$ increases from zero (shown by progressively darker lines, with $\Rey=0$ marked in blue for comparison) the peak of the pdf increases in amplitude.
In statistical terms, this means that the most common values of local heat flux become more common as shear is increased, up to $\Rey\approx 10^{3.5}$.
Due to the skewed nature of the distributions, these increasingly common values of heat flux are below the mean, and although the peak shifts slightly to the right as $\Rey$ increases, these parameters are associated with the drop in heat flux observed in figure \ref{fig:Nuss_comparison}.
This corresponds to the visualisation of figure \ref{fig:wall_planes}, where the streaks of low heat flux span a larger proportion of the boundary at $Ri=1$ (panels e,i) than at high $Ri$ (panels b,c).
Once $\Rey$ is sufficiently high, which in these cases is for $\Rey> 10^{3.5}$, the whole distribution shifts more significantly to higher $q_\theta$ as the transport becomes dominated by the pressure-driven shear flow.

At moderate $\Rey$, another key modification is to the tails of the distribution.
These are most clearly visualised in figure \ref{fig:local_flux_pdf}$(c,d)$, where the heat flux distributions are plotted on a logarithmic axis.
In this representation, it is clear that the right tails of large heat flux decay exponentially for $Gr=10^7$, and close to exponentially for $Gr=10^6$.
At both Grashof numbers, the tails are reduced as $\Rey$ increases relative to the natural convection case, showing that the probability of extreme local heat flux values is reduced due to the introduction of a mean horizontal flow.
Again, this is consistent with what was observed in the visualisation of figure \ref{fig:wall_planes}, where dark patches associated with large local heat flux became less prevalent as $\Rey$ increases.
In natural VC, \citet{pallares_turbulent_2010} used conditional sampling to show that these patches are most often associated with instantaneous flow reversals at the walls, which are likely a result of impacting plumes originating from the opposing wall.

\section{Conclusion and outlook \label{sec:conclusion}}

In this paper, we have investigated mixed convection in a differentially heated vertical channel subject to a horizontal pressure gradient, referred to as mixed VC, through direct numerical simulations.
By simulating the system across the parameter range $10^6 \leq Gr \leq 10^8$, ${1 \leq \Pran \leq 10}$, and ${10^{2.5} \leq \Rey \leq 10^4}$, we have explored the transition from natural convection to forced convection, characterised by the Richardson number $10^{-2} \leq Ri \leq 10^2$.
Across this parameter range, the response of the streamwise skin friction is identical to the response seen in mixed Rayleigh--Bénard (RB) convection, with a power-law $\Rey$-dependence for $C_f$ giving way to the Prandtl friction law \eqref{eq:Cf_Prandtl} at sufficiently high $\Rey$.
The presence of convection acts to increase the skin friction for a given $\Rey$, due to the thermal plumes emitted from the plates, with a collapse observed for all the data in the power-law regime of $C_f \sim Gr^{1/4} \Rey^{-\gamma}$ for both configurations across the entire range of $\Pran$.
For mixed VC, the streamwise momentum budgets show an enhanced impact of Reynolds stresses close to the wall at higher $Ri$, which modify the shape of the mean velocity profile.
The Reynolds stresses in the boundary layer are driven by the combined shear stress of the horizontal pressure-driven flow and the vertical convection flow.

Friction coefficients could also be obtained for the flow component associated with buoyancy driving.
For cases with significant buoyancy effects at $Ri > 1/4$, a reasonable collapse was found for the laminar-like scaling $C_f\sim Re_\delta^{-1}$ where $Re_\delta$ is the boundary layer Reynolds number of the convection flow.
The introduction of the horizontal pressure gradient leads to significant modification of the mean vertical velocity, with $Re_\delta$ increasing up to three times its value for natural convection, and a corresponding increase in the mean shear in the channel bulk for moderate $\Rey$.
For $Ri=O(1)$, the response of the Nusselt number $\Nu$ to the introduction of shear in mixed VC is also identical to that in mixed RB.
Before the flow undergoes a transition to a forced convection regime at high $\Rey$, the response can be expressed as $Nu/Nu_0 = f(\Rey/Re_0)$ where $Nu_0$ and $Re_0$ are the Nusselt number and Reynolds number associated with natural convection.
The data of both mixed convection systems collapse onto this single curve, which describes the drop in $\Nu$ as $\Rey$ increases.

The near identical quantitative response of mixed RB and mixed VC is observed despite striking qualitative differences between the two configurations in terms of large-scale flow organisation.
Whereas mixed RB features large convective rolls oriented along the streamwise axis, no such structures form in mixed VC except at low $Ri$ when the dynamics are solely dominated by the pressure gradient forcing.
The absence of a low wavenumber peak in the heat flux co-spectrum for mixed VC confirms that the advective heat flux across the domain is transported by eddies or plumes with a wide range of scales rather than by large coherent rolls.
Comparing the spectra from natural VC with mixed VC reveals that the organisation of plume structures with a horizontal scale of $\lambda \approx H$ is suppressed by the horizontal mean flow.
This is reflected also in the distribution of local heat flux at the boundaries, which shows that instantaneous events of extreme heat flux become less likely as $\Rey$ increases.
As the boundaries becomes dominated by streaky structures, the formation of localised plume structures is disrupted, consistent with the earlier interpretations of reduced heat flux in mixed RB \citep{domaradzki_direct_1988,scagliarini_heat-flux_2014}.

The striking agreement between the two channel configurations compared here, regardless of the gravity direction, opens up the question of how universal such a response in skin friction and heat flux is for other mixed convection systems.
The independence to the gravity direction suggests that our results may be applicable more generally to inclined layer convection, as studied by \cite{daniels_defect_2002} at low $Gr$, subject to a horizontal pressure gradient at comparable $\Gr$ and $\Rey$.
However, it is less clear how directly applicable the findings here are to a wall plume subject to a crossflow.
Such a scenario is relevant to environmental applications such as at a melting ice face subject to ambient ocean currents \citep{jackson_meltwater_2020}.
Whereas both boundaries impact the transport in mixed RB and mixed VC, a wall plume constantly entrains fluid from the ambient at its outer edge.
Recent work has provided an innovative way to simulate wall plumes at high $\Gr$, through studying temporally-growing boundary layers \citep{ke_turbulence_2023,wells_classical_2023}, and it would be fruitful to understand how the growth of these boundary layers is affected by the presence of a turbulent crossflow.
For ice-ocean interactions, the picture is further complicated through multicomponent transport \citep{howland_double-diffusive_2023} and the development of rough boundaries which are closely coupled to the flow structures \citep{couston_topography_2021,ravichandran_combined_2022}.

From a more theoretical standpoint, our current work has also emphasised how certain flow properties are noticeably modified under the transition from natural vertical convection to forced convection.
These include the sign of the near-wall Reynolds stress in figure \ref{fig:vmom_budget} and the distribution of the local wall heat flux in figure \ref{fig:local_flux_pdf}.
In both RB and VC, predictions have been made for a transition to the `ultimate' regime at sufficiently high buoyancy driving, where the boundary layers behave as turbulent boundary layers \citep{lohse_ultimate_2023}, although this regime has thus far been inaccessible to three-dimensional numerical simulation.
Analysis of the aforementioned statistical quantities in natural convection at high $\Gr$ may help in identifying key markers of such a transition in natural convection systems.


\backsection[Acknowledgements]{
    We are grateful to Emily Ching for fruitful discussions about natural vertical convection and to Olga Shishkina and Richard Stevens for insights into the response of mixed Rayleigh--Bénard convection.
}

\backsection[Funding]{
    The work of CJH was funded by the Max Planck Center for Complex Fluid Dynamics.
    The contribution of GSY towards this project has received funding from the European Research Council under the European Union’s Horizon 2020 research and innovation program (Grant No. 804283).
    We acknowledge PRACE for awarding us access to Irene at Très Grand Centre de Calcul (TGCC) du CEA, France (project 2021250115).
    For the extended domain simulations, the authors also gratefully acknowledge the Gauss Centre for Supercomputing e.V. (\url{www.gauss-centre.eu}) for funding this project (pr74sa) by providing computing time on the GCS Supercomputer SuperMUC-NG at Leibniz Supercomputing Centre (\url{www.lrz.de}).
    This work was also carried out on the Dutch national e-infrastructure with the support of SURF Cooperative.
}

\backsection[Declaration of interests]{The authors report no conflict of interest.}


\backsection[Author ORCIDs]{\\
    Christopher J. Howland \href{https://orcid.org/0000-0003-3686-9253}{https://orcid.org/0000-0003-3686-9253};\\
    Guru Sreevanshu Yerragolam \href{https://orcid.org/0000-0002-8928-2029}{https://orcid.org/0000-0002-8928-2029};\\
    Roberto Verzicco \href{https://orcid.org/0000-0002-2690-9998}{https://orcid.org/0000-0002-2690-9998};\\
    Detlef Lohse \href{https://orcid.org/0000-0003-4138-2255}{https://orcid.org/0000-0003-4138-2255}.
}


\appendix

\section{Alternative near-wall scaling for the streamwise Reynolds stress} \label{app:wall_scaling}

\begin{figure}
    \centering
    \includegraphics{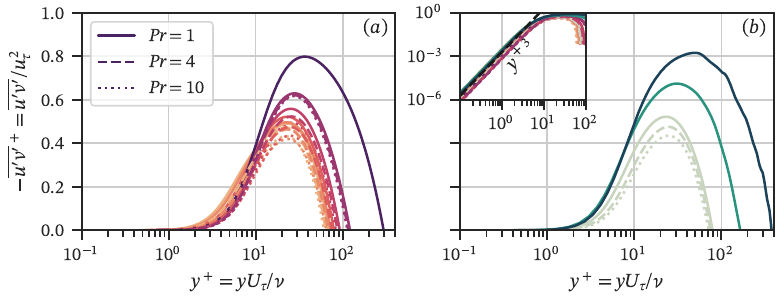}
    \caption{
        Wall-normal profiles of the streamwise Reynolds stress $\overline{uv}$ normalised by the streamwise friction velocity $u_\tau$ as previously shown in figure \ref{fig:uv_profiles}.
        Here, the wall-normal coordinate $y$ is scaled by the viscous wall unit $\nu/U_\tau$ computed from the \emph{total} wall shear stress $|\boldsymbol{\tau}|$.
        $(a)$ presents data at fixed $\Gr=10^6$ with $(b)$ at fixed $\Rey=10^3$, and the same colours and line styles as in figure \ref{fig:uv_profiles} are used to denote variation in $\Gr$, $\Rey$ and $\Pran$.
        As in figure \ref{fig:uv_profiles}, the inset panel presents both data on a double logarithmic axis.
        This figure is also available as an \href{https://cocalc.com/share/public_paths/c690a4a6317a519769e8c2b6c3e5c97a40c2338f/figure\%2015/figure\%2015.ipynb}{interactive JFM notebook}.
    }
    \label{fig:rescaled-uv-profiles}
\end{figure}

As shown in figure \ref{fig:uv_profiles}$(b,d)$, the near-wall profile of the streamwise Reynolds stress $\overline{uv}^+(y^+)$ requires an additional prefactor of $\Ri^{1/4}$ to collapse the majority of the cases studied here.
In \S\ref{sec:profiles}, we attribute this to the multiple components of shear at the wall that produce the Reynolds stress, which cannot be captured by the streamwise component $u_\tau$ alone.
This is confirmed in figure \ref{fig:rescaled-uv-profiles}, where we re-plot $\overline{uv}^+$ against a wall-normal coordinate scaled with the \emph{total} shear stress at the wall.
Specifically, we scale with the viscous wall unit $\nu/U_\tau$ where the total friction velocity $U_\tau$ satisfies
\begin{equation}
    U_\tau = \sqrt{\frac{|\boldsymbol{\tau}|}{\rho}} = \left(\frac{\sqrt{\tau_u^2 + \tau_w^2}}{\rho}\right)^{1/2} = \left( u_\tau^4 + w_\tau^4\right)^{1/4} . \label{eq:total_Utau}
\end{equation}
Compared to the results of figure \ref{fig:uv_profiles}, the previously outlying case of $\Rey=10^4$ (plotted as a dark purple line) now also shows a reasonable collapse with the rest of the data in the near wall region of figure \ref{fig:rescaled-uv-profiles}.
This result highlights the intricate nature of turbulence in mixed vertical convection, with different friction velocities needed to scale the two axes of figure \ref{fig:rescaled-uv-profiles} to describe the Reynolds stress.
Although the near-wall length scale is determined by the total shear stress in \eqref{eq:total_Utau}, the Reynolds stress must still satisfy the global balance \eqref{eq:mean_mom_int}, where $u_\tau$ is the relevant velocity scale.

\section{Simulation parameters \label{app:numerics}}

Table \ref{tab:params} details the physical and numerical input parameters used for the mixed vertical convection simulations.
Even at $\Pran=1$, we use a more refined grid for the temperature field since sharper gradients can emerge than in the velocity field due to the lack of a pressure gradient term in \eqref{eq:temp_evo} when compared with \eqref{eq:NSmom}.

In \S\ref{sec:spectra}, an additional set of simulations are discussed in which the periodic extent of the domain is tripled to an aspect ratio of $\Gamma=24$.
In that section, we compare RB configurations with gravity aligned normal to the boundary plates (in the negative $y$ direction) to VC configurations where gravity is aligned parallel to the plates (in the negative $z$ direction).
Table \ref{tab:spec_params} details the input parameters used for these additional simulations.
Note that a prefix of `P-' in the name of the simulation denotes the presence of a Poiseuille-like pressure gradient forcing.

\begin{table}
    \centering
    \begin{tabular}{ccccccccccc}
\toprule
$Gr$ & $Pr$ & $Re$ & $N_x=N_z$ & $N_y$ & $N_x^r$ & $N_y^r$ & $C_f^u$ & $Re_\delta$ & $C_f^w$ & $Nu$ \\
\midrule
$10^6$ & 1 & \SI{3.16e+02}{} & 512 & 192 & 1024 & 384 & \SI{7.84e-02}{} & 36.8 & \SI{1.66e-01}{} & 6.37 \\
$10^6$ & 1 & \SI{5.62e+02}{} & 512 & 192 & 1024 & 384 & \SI{4.46e-02}{} & 48.8 & \SI{1.30e-01}{} & 5.98 \\
$10^6$ & 1 & \SI{1.00e+03}{} & 512 & 192 & 1024 & 384 & \SI{2.59e-02}{} & 61.1 & \SI{1.00e-01}{} & 5.57 \\
$10^6$ & 1 & \SI{1.78e+03}{} & 512 & 192 & 1024 & 384 & \SI{1.57e-02}{} & 81.6 & \SI{7.62e-02}{} & 5.33 \\
$10^6$ & 1 & \SI{3.16e+03}{} & 768 & 192 & 1536 & 384 & \SI{1.06e-02}{} & 85.2 & \SI{6.89e-02}{} & 5.85 \\
$10^6$ & 1 & \SI{1.00e+04}{} & 1024 & 192 & 2048 & 384 & \SI{6.96e-03}{} & 43.2 & \SI{1.91e-01}{} & 12.01 \\
$10^6$ & 4 & \SI{3.16e+02}{} & 512 & 192 & 1024 & 384 & \SI{7.27e-02}{} & 29.9 & \SI{2.17e-01}{} & 9.39 \\
$10^6$ & 4 & \SI{5.62e+02}{} & 512 & 192 & 1024 & 384 & \SI{4.02e-02}{} & 41.3 & \SI{1.60e-01}{} & 8.69 \\
$10^6$ & 4 & \SI{1.00e+03}{} & 512 & 192 & 1024 & 384 & \SI{2.30e-02}{} & 59.6 & \SI{1.11e-01}{} & 8.15 \\
$10^6$ & 4 & \SI{1.78e+03}{} & 512 & 192 & 1024 & 384 & \SI{1.42e-02}{} & 74.6 & \SI{8.74e-02}{} & 8.39 \\
$10^6$ & 4 & \SI{3.16e+03}{} & 768 & 192 & 1536 & 384 & \SI{1.01e-02}{} & 61.3 & \SI{1.03e-01}{} & 10.46 \\
$10^6$ & 10 & \SI{3.16e+02}{} & 512 & 192 & 1536 & 576 & \SI{6.68e-02}{} & 28.7 & \SI{2.43e-01}{} & 11.48 \\
$10^6$ & 10 & \SI{5.62e+02}{} & 512 & 192 & 1536 & 576 & \SI{3.64e-02}{} & 39.7 & \SI{1.75e-01}{} & 10.73 \\
$10^6$ & 10 & \SI{1.00e+03}{} & 512 & 192 & 1536 & 576 & \SI{2.10e-02}{} & 58.2 & \SI{1.22e-01}{} & 10.38 \\
$10^6$ & 10 & \SI{1.78e+03}{} & 512 & 192 & 1536 & 576 & \SI{1.33e-02}{} & 69.3 & \SI{9.83e-02}{} & 11.16 \\
$10^6$ & 10 & \SI{3.16e+03}{} & 768 & 192 & 2304 & 576 & \SI{9.85e-03}{} & 45.3 & \SI{1.53e-01}{} & 15.04 \\
$10^7$ & 1 & \SI{3.16e+02}{} & 512 & 192 & 1024 & 384 & \SI{1.44e-01}{} & 90.3 & \SI{8.68e-02}{} & 13.74 \\
$10^7$ & 1 & \SI{1.00e+03}{} & 512 & 192 & 1024 & 384 & \SI{4.63e-02}{} & 99.8 & \SI{7.46e-02}{} & 13.14 \\
$10^7$ & 1 & \SI{1.78e+03}{} & 512 & 192 & 1024 & 384 & \SI{2.72e-02}{} & 118.9 & \SI{6.46e-02}{} & 12.47 \\
$10^7$ & 1 & \SI{3.16e+03}{} & 768 & 192 & 1536 & 384 & \SI{1.63e-02}{} & 136.7 & \SI{5.31e-02}{} & 11.85 \\
$10^7$ & 1 & \SI{5.62e+03}{} & 768 & 192 & 1536 & 384 & \SI{1.04e-02}{} & 157.9 & \SI{4.49e-02}{} & 12.06 \\
$10^7$ & 1 & \SI{1.00e+04}{} & 1024 & 256 & 2048 & 512 & \SI{7.52e-03}{} & 160.6 & \SI{4.53e-02}{} & 14.39 \\
$10^8$ & 1 & \SI{1.00e+03}{} & 1536 & 384 & 2304 & 512 & \SI{8.71e-02}{} & 198.0 & \SI{4.71e-02}{} & 28.14 \\
$10^8$ & 1 & \SI{3.16e+03}{} & 1536 & 384 & 2304 & 512 & \SI{2.90e-02}{} & 226.7 & \SI{4.17e-02}{} & 27.05 \\
$10^8$ & 1 & \SI{1.00e+04}{} & 1536 & 384 & 2304 & 512 & \SI{1.07e-02}{} & 266.6 & \SI{3.21e-02}{} & 25.82 \\
$10^6$ & 1 & \SI{0.00e+00}{} & 512 & 192 & 1024 & 384 & - & 31.7 & \SI{1.89e-01}{} & 6.62 \\
$10^6$ & 4 & \SI{0.00e+00}{} & 512 & 192 & 1024 & 384 & - & 23.7 & \SI{2.67e-01}{} & 10.09 \\
$10^6$ & 10 & \SI{0.00e+00}{} & 512 & 192 & 1536 & 576 & -  & 20.2 & \SI{3.42e-01}{} & 12.76 \\
$10^7$ & 1 & \SI{0.00e+00}{} & 512 & 192 & 1024 & 384 & - & 90.9 & \SI{8.59e-02}{} & 13.76 \\
$10^8$ & 1 & \SI{0.00e+00}{} & 1536 & 384 & 2304 & 576 & - & 181.3 & \SI{4.97e-02}{} & 28.24 \\
\bottomrule
\end{tabular}

    \caption{
        Physical control parameters: Grashof number $\Gr$, Prandtl number $\Pran$, Reynolds number $\Rey$; numerical grid parameters: number of grid points in the periodic ($x$, $z$) directions $N_x=N_z$ and wall-normal direction $N_y$ for the base grid and for the refined grid ($N_x^r=N_z^r$ and $N_y^r$); global response parameters: streamwise friction coefficient $C_f^u$, vertical boundary layer Reynolds number $Re_\delta$, vertical friction coefficient $C_f^w$, Nusselt number $\Nu$.
    }
\label{tab:params}
\end{table}

\begin{table}
    \centering
    \begin{tabular}{cccccccccccc}
\toprule
Name & $Gr$ & $Pr$ & $Re$ & $N_x=N_z$ & $N_y$ & $N_x^r$ & $N_y^r$ & $C_f^u$ & $Re_\delta$ & $C_f^w$ & $Nu$ \\
\midrule
RB & $10^7$ & 1 & \SI{0.00e+00}{} & 2304 & 192 & 4608 & 384 & - & 45.0 & \SI{1.65e-01}{} & 15.79 \\
VC & $10^7$ & 1 & \SI{0.00e+00}{} & 2304 & 192 & 4608 & 384 & - & 89.9 & \SI{8.74e-02}{} & 13.80 \\
P-RB & $10^7$ & 1 & \SI{3.16e+03}{} & 2304 & 192 & 4608 & 384 & \SI{1.47e-02}{} & 70.7 & \SI{9.91e-02}{} & 12.01 \\
P-VC & $10^7$ & 1 & \SI{3.16e+03}{} & 2304 & 192 & 4608 & 384 & \SI{1.62e-02}{} & 134.1 & \SI{5.45e-02}{} & 11.86 \\
\bottomrule
\end{tabular}
    \caption{
        Physical control parameters, numerical grid parameters, and global response parameters for the extended domain simulations with $\Gamma = 24$ discussed in \S\ref{sec:spectra}.
        The statistics for these simulations were only collected over 100 advective time units due to the increased computational cost of the larger domains and the spectra calculations.
    }
\label{tab:spec_params}
\end{table}

\section{Kinetic energy budgets}

For completeness, we finally provide an overview of the volume-averaged kinetic energy budgets for the pressure-driven mixed convection in a vertical channel.
All of the data presented here, along with additional second-order statistics may be found in the online JFM notebook.

Taking $y$ as the wall-normal coordinate, we can decompose the velocity field into a mean and a perturbation, where the mean is averaged in $x$, $z$ and $t$ (under the assumption of a statistically steady state):
\begin{align}
\boldsymbol{u} &= \overline{\boldsymbol{u}}(y) + \boldsymbol{u}'(x,y,z,t), &
\overline{\boldsymbol{u}} &= (\overline{u}(y), 0, \overline{w}(y)), &
\boldsymbol{u}' &= (u', v', w') .
\end{align}
The (volume-averaged) kinetic energy can thus be decomposed into mean and turbulent components
\begin{align}
\mathcal{K} &\equiv \frac{1}{2} \langle |\boldsymbol{u}|^2 \rangle = \frac{1}{2} \left( \langle \overline{u}^2 + \overline{w}^2 \rangle + \langle |\boldsymbol{u}'|^2 \rangle \right), &
\overline{\mathcal{K}} &\equiv \frac{1}{2} \langle \overline{u}^2 + \overline{w}^2 \rangle, &
\mathcal{K}' &\equiv \frac{1}{2} \langle |\boldsymbol{u}'|^2 \rangle .
\end{align}
From the momentum equation, we can then derive the budget equation for the total kinetic energy $\mathcal{K}$:
\begin{equation}
\underbrace{\frac{U u_\tau^2}{H/2}}_{\mathcal{I}} + \underbrace{g\alpha \langle w\theta \rangle}_{q} = \underbrace{\nu \left\langle \frac{\partial u_i}{\partial x_j} \frac{\partial u_i}{\partial x_j} \right\rangle}_{\varepsilon}
\end{equation}
Here, the energy input from the pressure gradient ($\mathcal{I}$) and the buoyancy flux ($q$) are balanced by viscous dissipation ($\varepsilon$).

\begin{figure}
    \centering
    \includegraphics{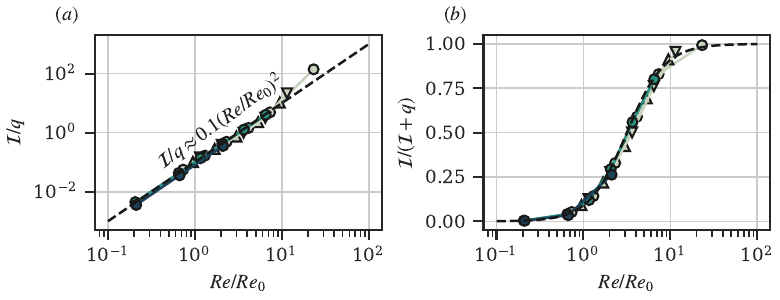}
    \caption{
        Components of the total kinetic energy budget in mixed vertical convection.
        $(a)$ Ratio of the energy input from the pressure gradient $\mathcal{I}$ to the energy input from the buoyancy $q$ as a function of $Re/Re_0$.
        $(b)$ Proportion of the total energy input produced by the pressure gradient.
        The dashed black line is an empirical fit to the data described by $\mathcal{I}/q=0.1(Re/Re_0)^2$.
        Colours and symbols denote variation in $Gr$ and $Pr$ following the figures of \S\ref{sec:global_response}.
    }
    \label{fig:total_KE_budget}
\end{figure}

Figure \ref{fig:total_KE_budget} presents how the relative importance of the two energy production terms changes as $Re$ is increased.
An excellent collapse is observed when the ratio $\mathcal{I}/q$ is plotted in terms of $Re/Re_0$, where $Re_0=W_0H/\nu$ is defined (as earlier) using the peak vertical velocity $W_0$ of the natural vertical convection flow at matching $Gr$ and $Pr$.
An empirical estimate of $\mathcal{I}/q\approx 0.1 (Re/Re_0)^2$ appears to describe the transition from buoyancy-dominant to pressure-dominant regimes although there is currently no theoretical justification for this relationship.
Although $\mathcal{I}$ can straightforwardly be related to the friction coefficient as $\mathcal{I}=C_f U^3/H$, there is no closed estimate for the vertical buoyancy flux $q$, even in natural vertical convection.

In a similar manner as for the total kinetic energy, we can also derive a budget equation for the mean kinetic energy $\overline{\mathcal{K}}$ as
\begin{equation}
\underbrace{\frac{U u_\tau^2}{H/2}}_{\mathcal{I}} + \underbrace{g\alpha \langle \overline{w}\overline{\theta} \rangle}_{\overline{q}} = \underbrace{\left\langle-\overline{v'\boldsymbol{u}'}\cdot\frac{\partial \overline{\boldsymbol{u}}}{\partial y}\right\rangle}_\mathcal{P} + \underbrace{\nu \left\langle \left| \frac{\partial \overline{\boldsymbol{u}}}{\partial y}\right|^2 \right\rangle}_{\overline{\varepsilon}}
\end{equation}
where the shear production $\mathcal{P}$ converts mean kinetic energy to turbulent kinetic energy.
Note that the two (horizontal and vertical) components of the mean kinetic energy can be completely decoupled, such that we can derive separate budgets for those quantities:
\begin{align}
\underbrace{\frac{U u_\tau^2}{H/2}}_{\mathcal{I}} &= \underbrace{\left\langle-\overline{v'u'}\frac{\partial \overline{u}}{\partial y}\right\rangle}_{\mathcal{P}_u} + \underbrace{\nu \left\langle \left| \frac{\partial \overline{u}}{\partial y}\right|^2 \right\rangle}_{\overline{\varepsilon}_u} &
\underbrace{g\alpha \langle \overline{w}\overline{\theta} \rangle}_{\overline{q}} &= \underbrace{\left\langle-\overline{v'w'}\frac{\partial \overline{w}}{\partial y}\right\rangle}_{\mathcal{P}_w} + \underbrace{\nu \left\langle \left| \frac{\partial \overline{w}}{\partial y}\right|^2 \right\rangle}_{\overline{\varepsilon}_w}
\end{align}

\begin{figure}
    \centering
    \includegraphics{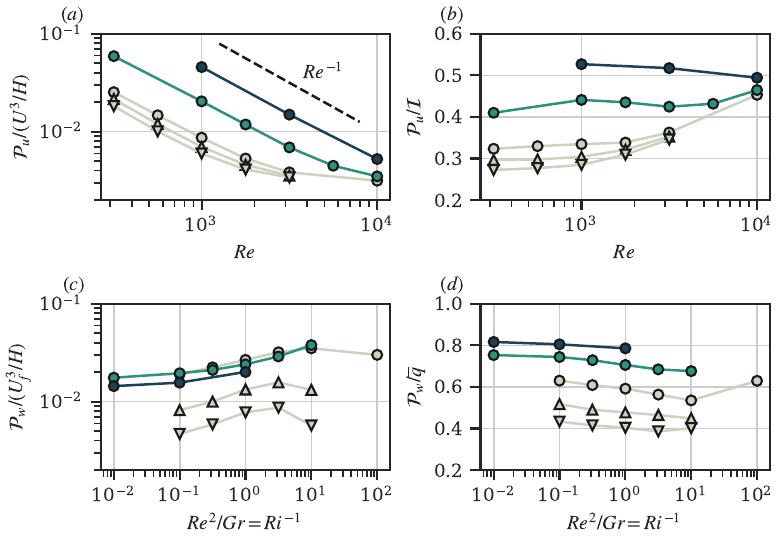}
    \caption{
        Budget contributions for the horizontal $(a,b)$ and vertical $(c,d)$ components of the mean kinetic energy.
        The horizontal component of shear production $\mathcal{P}_u$ is normalised by $(a)$ the bulk velocity scaling $U^3/H$ and $(b)$ the total energy injection due to the pressure gradient $\mathcal{I}$.
        The vertical component of shear production $\mathcal{P}_w$ is normalised by $(C)$ the free-fall velocity scaling $U_f^3/H$ and $(d)$ 
    }
    \label{fig:mean_KE_budget}
\end{figure}
The relative energy transfers in each of these budgets are shown in figure \ref{fig:mean_KE_budget}.
For the horizontal component of mean kinetic energy presented in panels $(a,b)$ we find that the proportion of energy transferred to TKE $\mathcal{P}_u/\mathcal{I}$ is roughly constant at low $Re$ for a given $Gr$, but increases at higher $Re$, with the trend suggesting independence of $Gr$ at sufficiently high $Re$.
This is reminiscent of the behaviour of $C_f$ in figure \ref{fig:Cf_Re_horiz}, perhaps unsurprisingly since $C_f=\mathcal{I}/(U^3/H)$.
Such a trend is not evident in the vertical component of the mean kinetic energy shown in panels $(c,d)$.
The vertical shear production $\mathcal{P}_w$ exhibits non-monotonic dependence on $Re$, and a reasonable collapse with $Ri$ for the $Pr=1$ data, but significant dependence on $Pr$ otherwise.
Larger $Pr$ and lower $Gr$ lead to a larger proportion of the energy supplied by the mean buoyancy flux $\overline{q}$ being directly dissipated by viscosity (through $\overline{\varepsilon}_w$).

Taking the difference of the total and mean kinetic energy budgets constructs the turbulent kinetic energy budget for our system, which reads
\begin{equation}
\underbrace{-\overline{v'\boldsymbol{u}'}\frac{\partial \overline{\boldsymbol{u}}}{\partial y}}_\mathcal{P} + \underbrace{g\alpha \langle w' \theta' \rangle}_{q'} = \underbrace{\nu \left\langle \frac{\partial u_i'}{\partial x_j} \frac{\partial u_i'}{\partial x_j} \right\rangle}_{\varepsilon'}
\end{equation}
\begin{figure}
    \centering
    \includegraphics{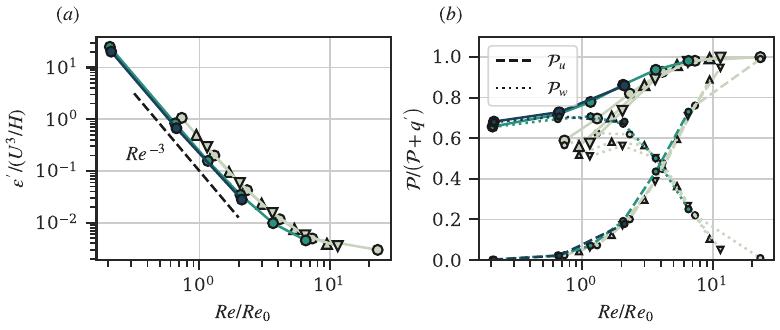}
    \caption{
        Budget contributions for the turbulent kinetic energy.
        $(a)$ TKE dissipation rate $\varepsilon'$ normalised by $U^3/H$ as a function of $Re/Re_0$;
        $(b)$ Proportion of TKE produced by shear, decomposed into horizontal (dashed) and vertical (dotted) components.
        Symbols for the decomposed terms are made smaller than those for the total shear production.
    }
    \label{fig:TKE_budget}
\end{figure}
Figure \ref{fig:TKE_budget}$(a)$ shows that the TKE dissipation rate $\varepsilon'$ data collapses as a function of $Re/Re_0$ when normalised by $U^3/H$.
For $Re\leq Re_0$, a $Re^{-3}$ scaling implies that $\epsilon'$ is independent of $U$, whereas at higher $Re/Re_0$ the slope flattens out, potentially consistent with a finite value of dissipation as $Re\rightarrow \infty$.
The remainder of the energy budget terms, shown in figure \ref{fig:TKE_budget}$(b)$ also show a good collapse with $Re/Re_0$, highlighting the crossover from the natural convection limit where the vertical component of shear production $\mathcal{P}_w$ and the turbulent buoyancy flux $q'$ balance the dissipation to the turbulent shear flow limit where the TKE production is dominated by the horizontal shear production $\mathcal{P}_u$.
The collapse of the data in the transitional values of $Re/Re_0$ is somewhat surprising given that in the natural convection limit, the ratios of the budget terms varies significantly with $Gr$ and $Pr$ \citep{howland_boundary_2022}.

\bibliographystyle{jfm}
\bibliography{mixedVC}

\end{document}